\documentclass[useAMS,usenatbib,twocolumn]{mn2e}
\usepackage{graphicx}
\usepackage{subfigure}

\newcommand{\mpt}{\mathrm{.}}

\newcommand{\Vec}[1]{ \mbox{\boldmath$ #1 $} }

\newcommand{\apj}{ApJ}
\newcommand{\apjs}{ApJS}
\newcommand{\mnras}{MNRAS}

\newcommand{\aj}{AJ}
\newcommand{\aap}{A\&A}

\title[Non-parametric inversion of lensing systems with few images]
      {Non-parametric inversion of gravitational lensing systems with
       few images using a multi-objective genetic algorithm}
\author[J. Liesenborgs, S. De Rijcke, H. Dejonghe and P. Bekaert]
{J. Liesenborgs$^1$\thanks{Corresponding author:
jori.liesenborgs@uhasselt.be}, S. De Rijcke$^2$\thanks{Postdoctoral
Fellow of the Fund for Scientific Research - Flanders
(Belgium)(F.W.O)}, H. Dejonghe$^2$ and P. Bekaert$^1$\\ $^1$
Expertisecentrum voor Digitale Media, Universiteit Hasselt,
Wetenschapspark 2, B-3590, Diepenbeek, Belgium \\ $^2$ Sterrenkundig
Observatorium, Universiteit Gent, Krijgslaan 281, S9, B-9000, Gent,
Belgium}

\begin{document}
	
\date{} 
		
\pagerange{\pageref{firstpage}--\pageref{lastpage}} \pubyear{2007}
		
\maketitle \label{firstpage} 
	
\begin{abstract} 
Galaxies acting as gravitational lenses are surrounded by, at most, a
handful of images. This apparent paucity of information forces one to
make the best possible use of what information is available to invert
the lens system. In this paper, we explore the use of a genetic
algorithm to invert in a non-parametric way strong lensing systems
containing only a small number of images. Perhaps the most important
conclusion of this paper is that it is possible to infer the mass
distribution of such gravitational lens systems using a non-parametric
technique. We show that including information about the null space
(i.e. the region where no images are found) is prerequisite to avoid
the prediction of a large number of spurious images, and to reliably
reconstruct the lens mass density. While the total mass of the lens is
usually constrained within a few percent, the fidelity of the 
reconstruction of the lens mass distribution depends on the number and 
position of the images. The technique employed to include null space 
information can be extended in a straightforward way to add additional 
constraints, such as weak lensing data or time delay information.
\end{abstract}
		
		
\section{Introduction}
		
The deflection of light from a distant source by the gravitational
influence of an intermediate object gives rise to so-called
gravitational lensing systems. Images of a source not only provide
information about said source, but also encode properties about
the mass of the object responsible for the deflection: the
gravitational lens.

Gravitational lens inversion, i.e. the determination of the mass
density of the lens based on observed image properties, is the focus
of this article. Lens inversion procedures are often divided into two
categories: parametric and non-parametric methods. Parametric
techniques approximate the mass distribution of the lens
by a function that is characterized by a small number of
parameters. They then optimize these parameters to provide the best
possible fit to the observed data. Several such algorithms have been
proposed (e.g. {\sc lensclean} \citep{LensCLEAN}), and several
software packages are publicly available (e.g. {\sc gravlens}
\citep{KeetonGravlens} and {\sc lenstool} \citep{KneibLenstool}).
Non-parametric inversion methods try
to avoid this restriction, for example by pixellizing the mass
distribution (e.g. \citet{Saha}), pixellizing the lens potential
(e.g. \citet{Bradac}, \citet{Koopmans}) or by dynamically adjusting
the number of basis functions used (e.g. \citet{Diego}).
		
In a previous article \citep{Liesenborgs}, we described a
non-parametric inversion routine, intended for strong lensing systems
containing many multiply imaged sources. Using a genetic algorithm, we
showed through simulations that the mass density of the lens can be
reliably retrieved and that the reconstructed sources closely resemble
their original counterparts. While such data is currently available
\citep{Broadhurst} and more of these systems are likely to be
discovered in the future, they are currently far outnumbered by
systems containing a handful of images of only one or a few sources.

For this reason, we felt the need to investigate the performance of
the original procedure and the required modifications when confronted
with a few-sources situation.  As in the previous work, simulations
are used to determine what can be inferred about the mass density when
all the necessary information is available with great accuracy,
thereby placing fundamental limits on the amount of information which
can be extracted from such a lensing scenario.  Again, we will try to
avoid imposing any prior on the mass density of the lens: only the
information contained in the observed images is used.

The situation being studied here, bears resemblance to the work of
\citet{Abdelsalam}: there, based on the image locations of a few
sources, a pixellized mass distribution of a cluster lens was
reconstructed. However, while their algorithm uses one position per
image, the method presented here makes use of the complete
images. Another important difference is that the method described in
that article searches for the mass distribution which follows the
light map as closely as possible. Below, no information about the
light map of the gravitational lens will be used.

The following section contains a brief review of the lensing
formalism, genetic algorithms and the original inversion
technique. Section \ref{sec:ext} illustrates the difficulties
encountered with the procedure and describes which measures can be
taken to overcome them. After illustrating the effect of these
modifications in sections \ref{sec:sim} and \ref{sec:sim2}, our
conclusions are summarized in section \ref{sec:conc}.
		
\section{Review}

\subsection{Lensing formalism}

To first post-Newtonian order, the gravitational lens effect can be
well described by the so-called lens equation, describing a mapping
$\Vec{\beta}(\Vec{\theta})$ from the image plane onto the source plane.
Our focus is on the strong 
lensing effect, in which multiple images of a source are generated. 
It is important to note that when these images are projected back 
onto the source plane using the lens equation, they must coincide 
since they are images of the same source. A thorough review of the
gravitational lensing formalism can be found in \citet{SchneiderBook}.
			
\subsection{Genetic algorithms}

Genetic algorithms implement a problem-solving strategy based on the
principle of natural selection, present in biological systems. The
procedure starts with a `population' of trial solutions, stored in an
encoded form which is often referred to as the genome. Each member of
the population is then evaluated and a fitness measure is
assigned. Using the current population, a new one is created by
combining genomes -- mimicking sexual reproduction -- or by copying
them -- mimicking asexual reproduction. When selecting genomes in this
process, a key ingredient of the genetic algorithm is to apply some
form of selection pressure: genomes which are deemed more fit, should
have a higher probability of creating offspring. This can be achieved
by ranking the genomes according to their fitness measures and by
assigning a higher selection probability to the genomes with a more
favorable rank. After introducing some random mutations to ensure
genetic diversity, the newly created population replaces the old one
and the procedure is repeated until a stopping criterion is
fulfilled. In some cases one can choose to apply `elitism' and copy
the best genome in a generation directly to the next generation.

Research into this kind of evolutionary optimization was pioneered by
\citet{Holland} and the type of genetic algorithm introduced there is
now referred to as the canonical genetic algorithm. Currently, genetic
algorithms are available in many forms and are used to solve a wide
variety of (often high-dimensional) problems, including the automatic
generation of computer algorithms (e.g. \citet{Koza}). For an
excellent introduction to the use of genetic algorithms in an
astrophysical context, the interested reader is referred to
\citet{Charbonneau}.

\subsection{Inversion method}
			
\begin{figure}
\centering \includegraphics[width=0.48\textwidth]{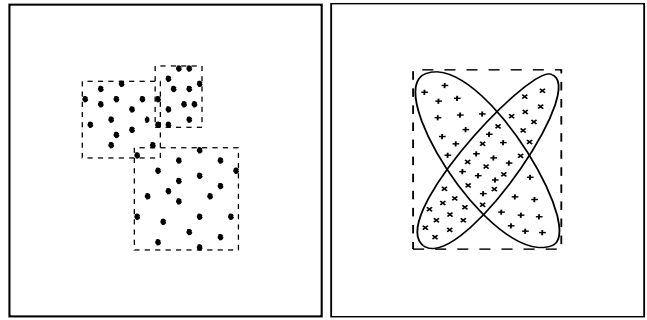}
\caption{Left panel:~back-projected images are enclosed by rectangles
of which the sides are parallel to the coordinate axes. The genetic
algorithm produces mass distributions which cause these rectangles to
overlap increasingly well. Right panel:~when different images of an
elliptical source are projected back onto the source plane, the
situation shown has a good fitness value, even though the images do
not overlap well. This problem is easily resolved by allowing the
enclosing rectangles to be rotated.}
\label{fig:backproj}
\end{figure}

The inversion procedure requires the user to define a square region in
which the strong lensing mass will be reconstructed. This region is
subdivided in a uniform grid of cells.  Each of the grid cells is
assigned a basis function, more specifically a projected Plummer
sphere \citep{Plummer}. Writing the mass density distribution of the
gravitational lens as a weighted sum of these basis functions, the
gravitational lens effect is described by the following lens equation:
\begin{equation} 
\Vec{\beta}(\Vec{\theta}) = \Vec{\theta} - \frac{D_{ds}}{D_s D_d} \frac{4 G}{c^2} \sum_{i = 1}^N \frac{\Vec{\theta}-\Vec{\theta}_{s,i}}{|\Vec{\theta}-\Vec{\theta}_{s,i}|^2+\theta_{P,i}^2} M_i \mpt
\label{eq:lenseqnmplum}
\end{equation}
Here, $\Vec{\theta}_{s,i}$ is the center of a basis function, $M_i$ is
its total mass and $\theta_{P,i}$ a measure of its width.  A genetic
algorithm is then applied to determine the weights $M_i$ of each basis
function and the resulting mass distribution is used to create a new
grid, in which regions containing a larger fraction of the mass will
be subdivided further. A Plummer basis function is assigned to each
new grid cell and the genetic algorithm again determines the
appropriate weights. This procedure can be repeated a number of times,
depending on the desired level of detail. 

The genetic algorithm is
clearly at the heart of this inversion method. For details about
genome encoding, reproduction and mutation, the reader is referred to
\citet{Liesenborgs}. 
Here, we will focus on the fitness evaluation since this determines
the evolution of the algorithm towards a solution. The input of the
entire procedure consists of the images associated with each
source. Using the Plummer weights $M_i$ stored in a genome, equation
(\ref{eq:lenseqnmplum}) is used to project these images back onto
their source planes. As was stated earlier, the correct lens equation
generates overlapping images for each source, and it is precisely this
criterion that is applied to determine the fitness of a genome. The
left panel of Fig.~\ref{fig:backproj} illustrates how the degree of
overlap is measured. Each back-projected image is surrounded by a
rectangle of which the sides are parallel to the coordinate
axes. If the back-projected images of a source overlap perfectly, so 
will the rectangles that surround each image. For this reason, the degree 
to which the images overlap will be approximated by the overlap of the
associated rectangles. This can easily be calculated: two rectangles
will coincide if their corners coincide, which suggests that the 
separation between corresponding corners can be used to measure the
amount of overlap. To assign a fitness value, corresponding corners 
of the rectangles are connected with imaginary springs and the potential 
energy of the situation is calculated. Note that in calculating distances 
between corners, the average size of the rectangles is used as the length 
scale. This makes sure that the overlap is measured relative to the
size of the source. The same
procedure is repeated for the other sources and the sum of all these
potential energies is used as the fitness measure of the genome. Lower
values indicate a better overlap of the back-projected images and
therefore correspond to genomes which are deemed more fit. Several
runs of this inversion routine typically generate solutions which
differ somewhat. This is a result of the inherent randomness present
in genetic algorithms which will manifest itself especially when
the grid is refined further. Averaging a number of individual
solutions yields a final mass density which is far more smooth than
any individual solution and inspecting the standard deviation provides
information about the amount of disagreement between the generated
solutions. This way, the standard deviation of individual solutions 
yields an indication of the uncertainty in the reconstructed mass 
density. Note that this is only an indication, as it does not include 
systematic effects.
	
Tests with simulations indicate that when many multiply imaged sources
are available, this procedure not only creates an excellent map of the
lensing mass, but also retrieves the sizes and positions of the
sources remarkably well. No prior assumptions are made regarding the
mass density or the sources, and by construction, no negative mass
densities will arise. The radial images are used in an effective way,
i.e. the relative weight of a given image in the fitness value does
not depend on its surface area, as in image fitting methods. This is
an important feature if one is interested in obtaining the best
possible estimate of the central mass density of the lens.
			
Like other non-parametric inversion algorithms, in reality there are
simply a very large number of parameters: the weights of a large
amount of basis functions. Plummer basis functions were chosen
mainly for simplicity, but also because earlier tests indicated
that a wide variety of mass distributions can be approximated by
a sum of Plummer basis functions. Similar inversion results can
be obtained using Gaussian basis functions; square basis
functions appear to be less suitable as they quickly cause complex
caustic structures. 

\section{A first test and algorithm extensions} \label{sec:ext}

\begin{figure*}
\centering
\subfigure{\includegraphics[origin=c,angle=-90,width=0.32\textwidth]{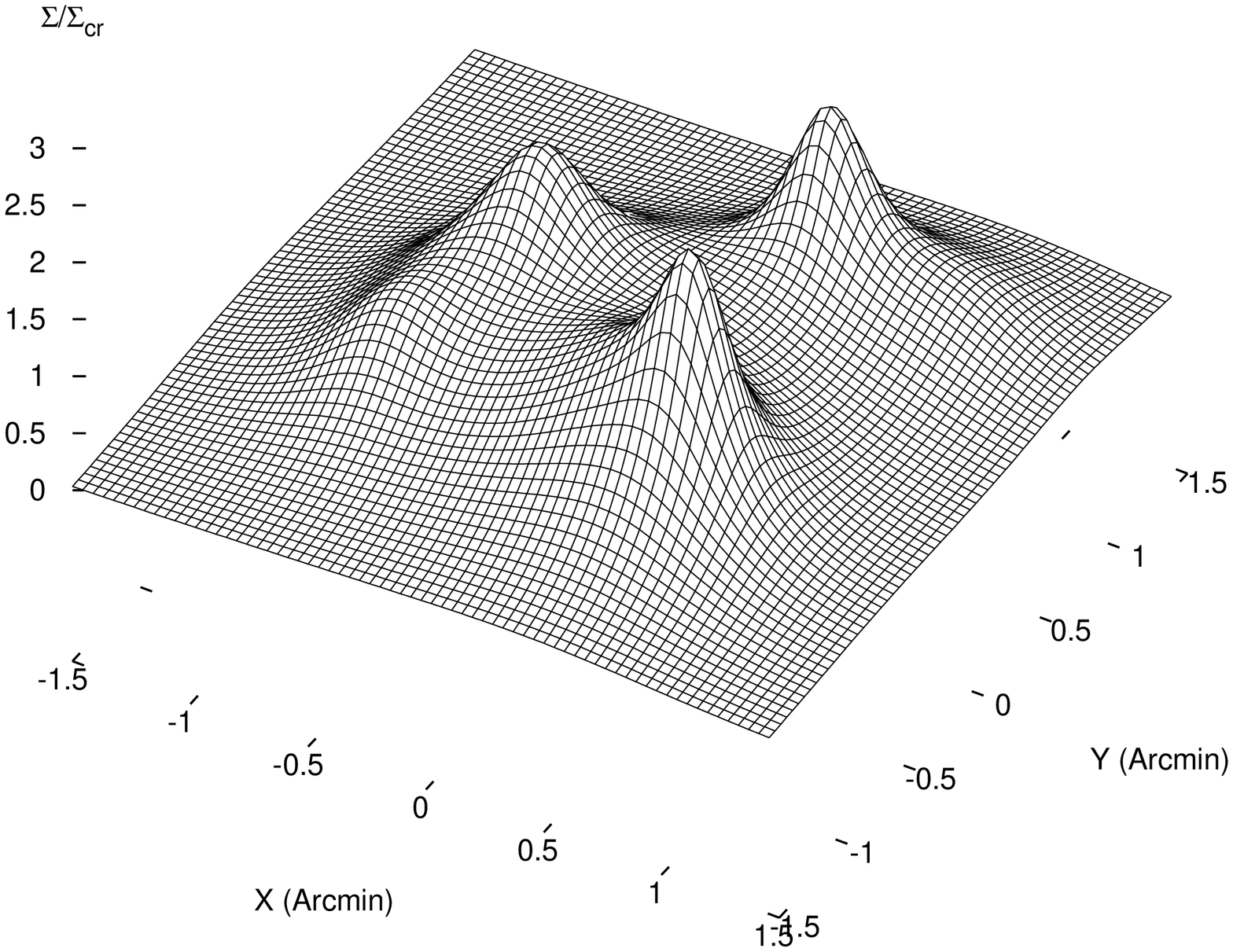}}
\subfigure{\includegraphics[width=0.32\textwidth]{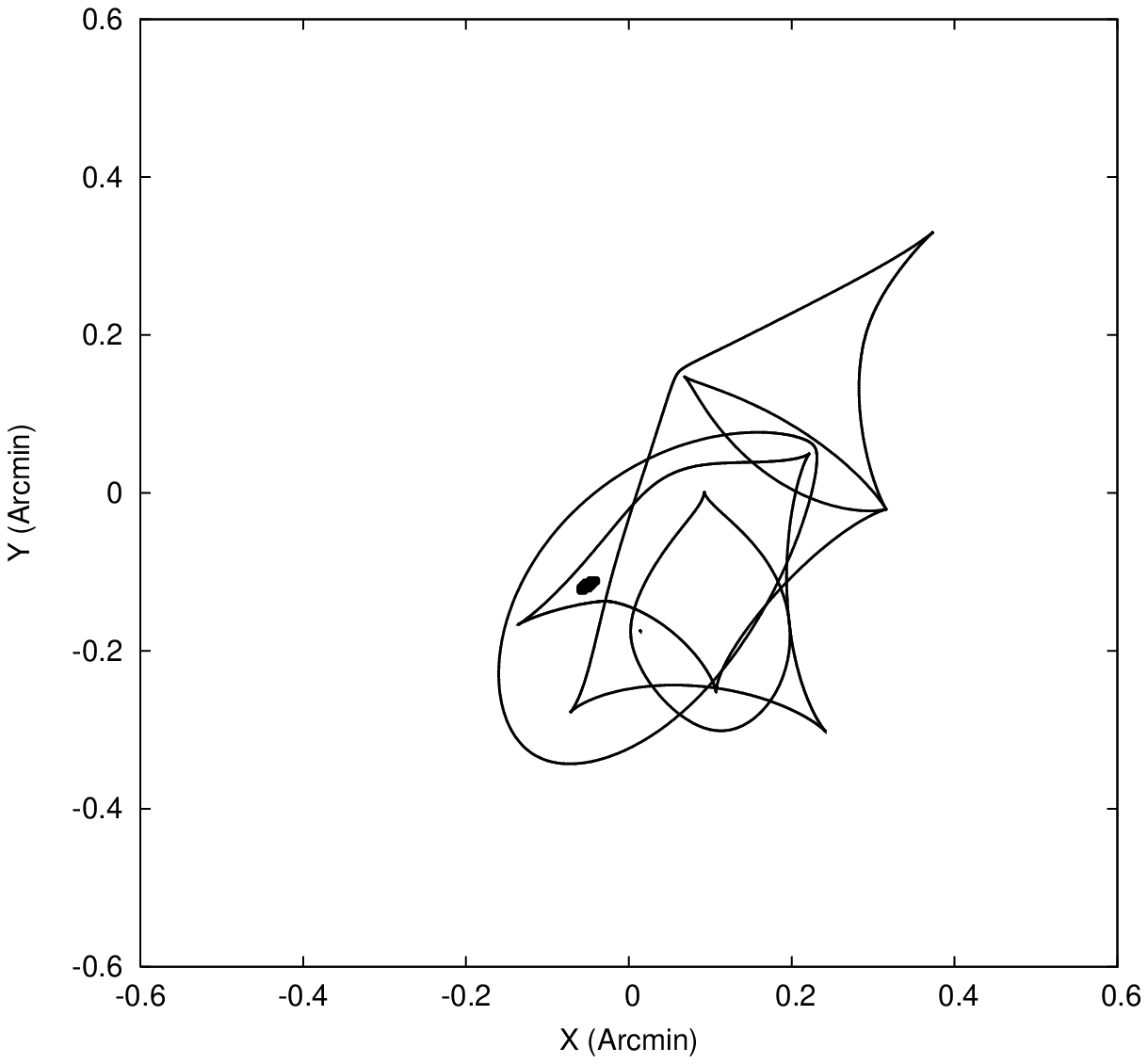}}
\subfigure{\includegraphics[width=0.32\textwidth]{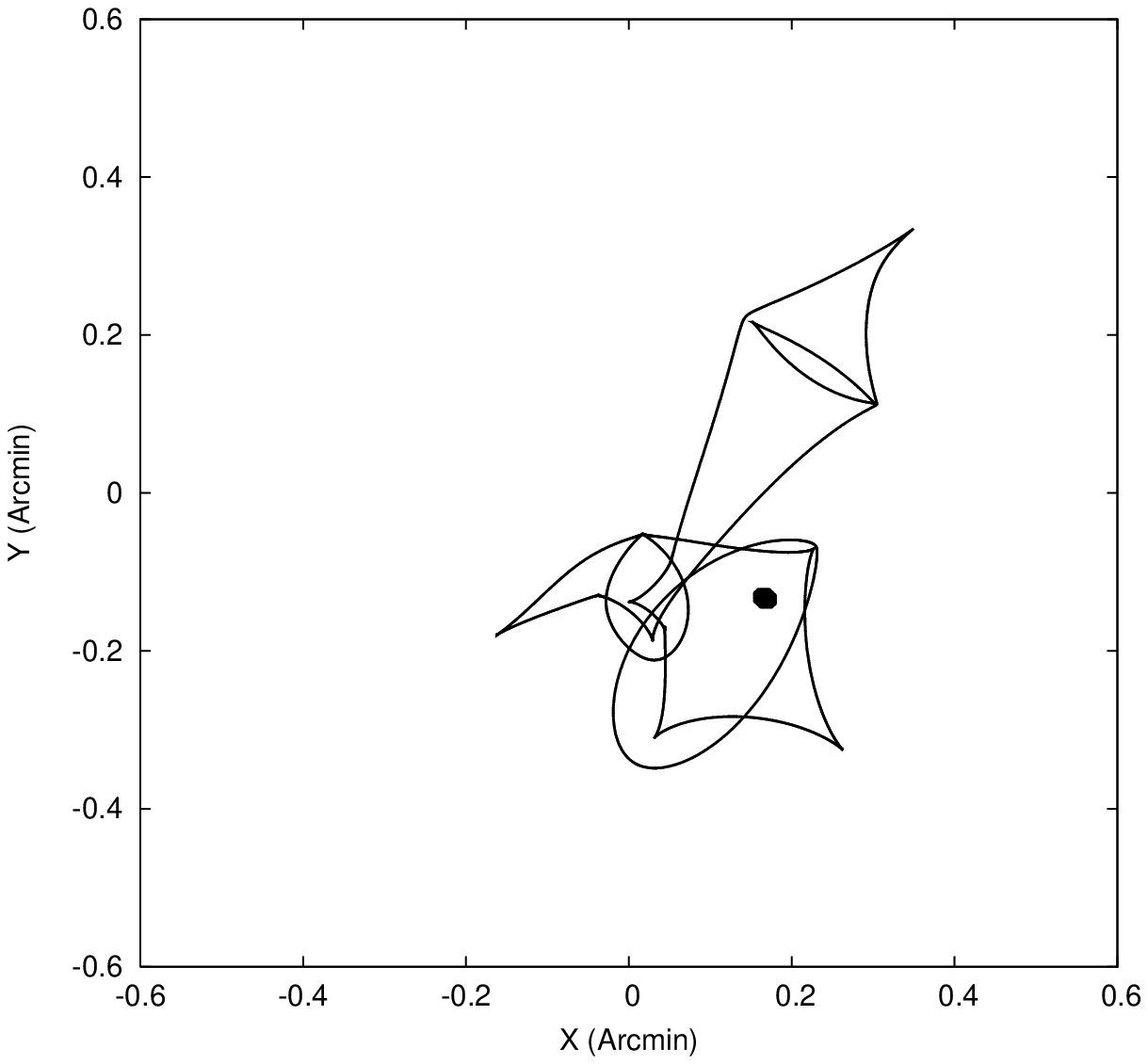}}
\caption{Left panel:~mass distribution of the input lens (placed
at redshift $0.45$) used in the first simulation. The value of $\Sigma_{cr}$
was evaluated at redshift $2.5$.
Middle panel and right panel:~the two sources, placed at redshifts
$2.5$ and $1.5$ respectively, used in the first simulation, together
with the caustic structures at these redshifts.}
\label{fig:reallens}
\end{figure*}

The inversion technique described above is intended to be used when
many multiply imaged sources are available to constrain the mass
distribution of the lens. To illustrate the problems one encounters in
a situation where the lensing mass is poorly sampled, we use the
sources and the lens mass density shown in Fig.~\ref{fig:reallens}. The
sources have an elliptical shape and are positioned at $z=2.5$ and
$z=1.5$ respectively; the lensing mass is located at $z=0.45$. 
In this article, we use a standard CDM cosmology with a matter density 
$\Omega=1$ and a Hubble parameter $H_0 = 70$~km~s$^{-1}$~Mpc$^{-1}$ 
for simplicity. As can be derived from the
caustic structures, each source produces five images, the positions of 
which are displayed in the left panel of Fig.~\ref{fig:realimages}. 
The critical lines corresponding to both redshifts are also indicated
in this figure.

\subsection{Difficulties}

\begin{figure*}
\centering
\subfigure{\includegraphics[width=0.48\textwidth]{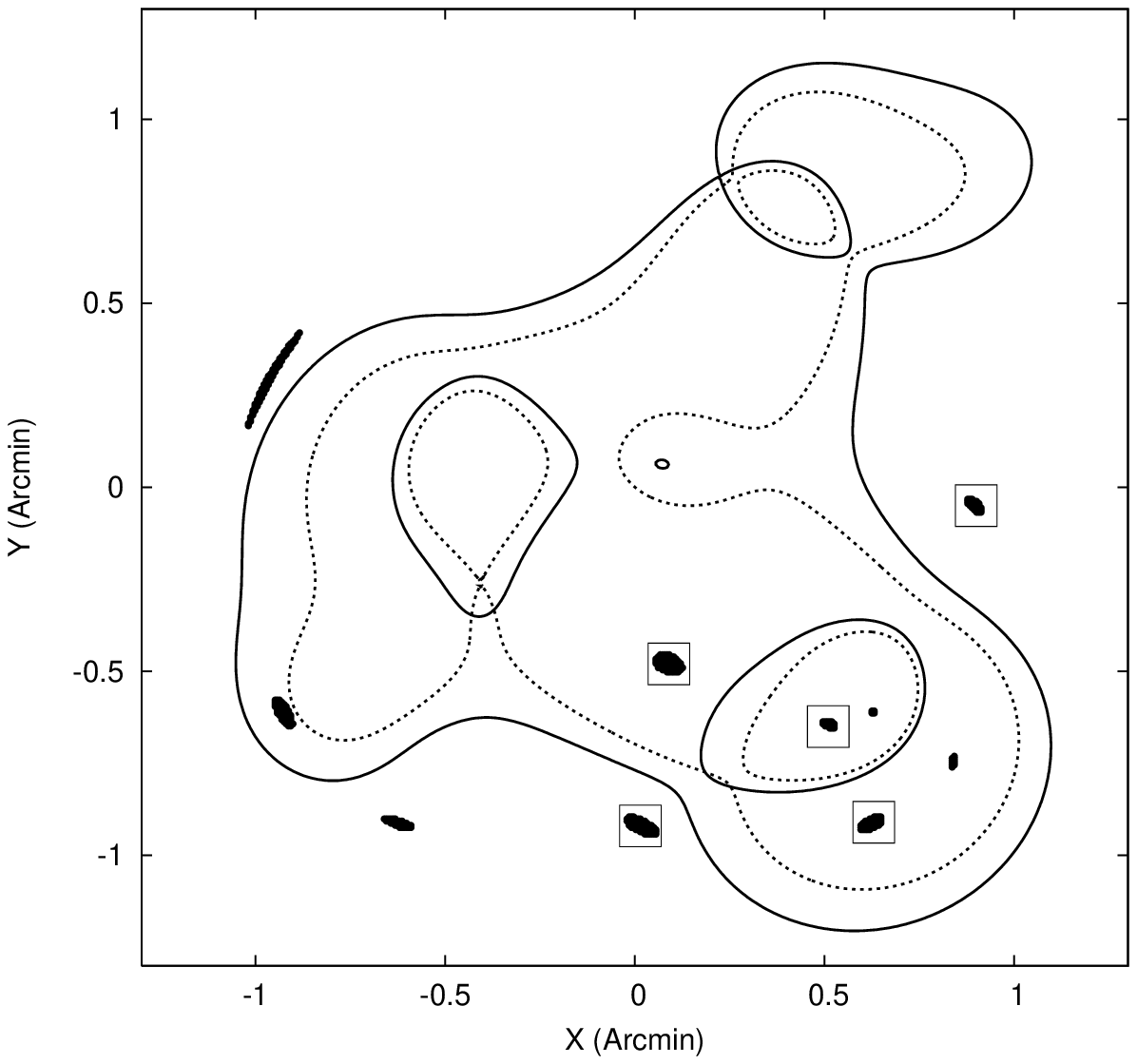}}
\subfigure{\includegraphics[width=0.48\textwidth]{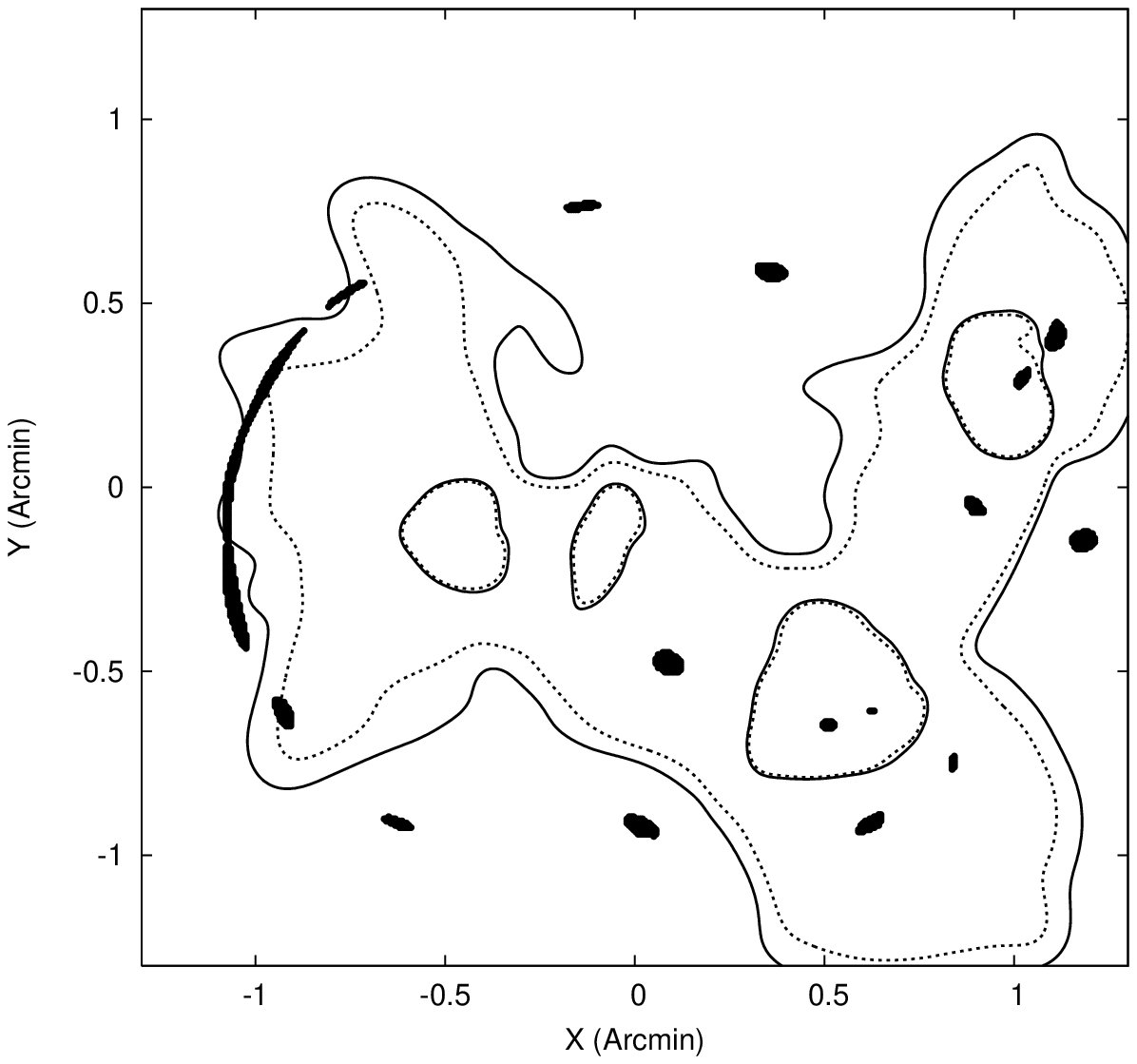}}
\caption{Left panel:~the lens and sources of Fig.~\ref{fig:reallens}
cause ten images to be generated. The critical lines of the first
source are shown as solid lines, those of the second source as
dashed lines. The images of the second source are enclosed by small
squares.
Right panel:~the original algorithm can easily
produce solutions which cause the back-projected images to overlap
well. However, the reconstructed sources and lens in general
cause many undesired images to be produced, as is illustrated
in this figure.}
\label{fig:realimages}
\end{figure*}

When the procedure outlined above is applied to the images of
Fig.~\ref{fig:realimages} (left panel), some undesired effects can be
observed. The first one is demonstrated in the right panel of
Fig.~\ref{fig:backproj}. Because the rectangles used to determine the
degree of overlap in the back-projected images are aligned with the
coordinate axes, crossing elliptical shapes will be interpreted as
overlapping images. Due to the large number of multiply imaged sources
in the original study, and hence the large number of independent
constraints, this problem did not manifest itself before. The solution
to the problem is clear: the rectangles which are placed around the
back-projected images should be rotated until they are aligned with
their respective images. This can be done in an efficient way using
the `rotating calipers' algorithm \citep{Shamos}.

\begin{figure}
\centering
\includegraphics[angle=0,width=0.48\textwidth]{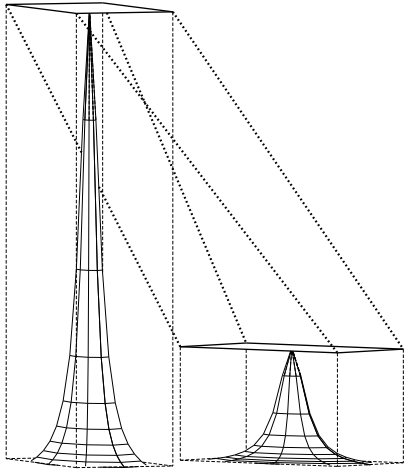}
\caption{Back-projected images are surrounded by rotated rectangles
and placed at a height corresponding to the maximal brightness values
of the images.  Corresponding corners are connected with virtual
springs (indicated by thick dotted lines) and the potential energy of
the situation is used as the fitness measure. This way, reconstructed
sources will also overlap in the brightness domain.}
\label{fig:heights}
\end{figure}

Another problem that surfaced due to the limited number of
constraints, is non-overlapping brightness distributions of
back-projected images. However, the fitness measure can easily be
modified to correct this behavior: each rectangle is placed at a
height corresponding to the maximal brightness value of the
corresponding image. The average height is used as length scale in
this direction and the fitness value is calculated as before, now
using three-dimensional virtual springs instead of two-dimensional
ones. The situation is illustrated in Fig.~\ref{fig:heights}. More
elaborate schemes can easily be implemented to ensure a suitable
overlap of the back-projected images, at the cost of increasing the
computer work load, but this will suffice for our goals. When these
extensions are added to the original genetic algorithm, it is able to
create solutions which produce overlapping images when projected back
onto the source plane, both in the position and brightness domains.

Unfortunately, when comparing the images predicted by the best lens
solution with the observed ones, a more serious problem presented
itself.  Because of the non-parametric nature of the reconstruction
and the lack of firm constraints, the reconstructed lens mass
distribution can contain many small-scale fluctuations that
generate a host of images other than the observed ones. The right
panel of Fig.~\ref{fig:realimages} illustrates this
undesired behavior.
		
\subsection{Using the null space}

One possible solution to avoid this multitude of spurious images, is
to impose some kind of prior on the lensing mass. For example, one
could modify the fitness criterion to favor smoother mass
distributions. Undoubtedly, this would reduce the number of extra
images since smoother mass distributions have less complex caustic
structures. However, this goes against the spirit of our endeavor,
which is to find out how much information can be retrieved
non-parametrically from a few-image lens system.
		
In order to avoid introducing this kind of bias in the generated
solutions, a different approach is used. Note that there is still much
information available that has not yet been used. Points in the image
plane which are not part of an image of the source, i.e.  the null
space, provide additional constraints: if they are projected back onto
the source plane using the correct lens equation, none of these points
will lie inside the source area. Otherwise, an image would be visible
at that specific location.  Using the null space was already suggested
in \citet{Diego}, but there, only null space points adjacent to the
images were used. This can avoid the acceptance of solutions that
produce images larger than the observed ones, but it obviously fails
to avoid the extra images.

To incorporate the information in the null space, two schemes have
been explored. In the first one, a regular grid of null space points
is generated and the points which fall inside an input image are
removed. To avoid generating images which lie relatively far from the
other images, and thereby generating more mass than is necessary, the
area in which the null space points are generated should be chosen
large enough, i.e. somewhat larger than the area of the images
themselves. An appropriate size can easily be discovered after a few
attempts, but 10-20\% larger is usually adequate. After projecting 
the images back onto the source plane,
their envelope is calculated. This is the smallest convex polygon
enclosing all the back-projected points, and is used as the current
estimate of the source shape. Next, each point in the null space is
projected back onto the source plane and the number of points which
lie inside the reconstructed source are counted.  Clearly, one would
like this count to be as low as possible, since our objective is to
remove the extra images. Note that by only considering a discrete
number of points in the null space, it is possible that small images
are generated which lie in between null space points.

Instead of simply using points, one can also divide the null space
into a number of small triangles, similar to the approach in
\citet{Blandford}. When these are projected back
onto the source plane, for each triangle the amount of overlap with 
the reconstructed source is calculated. The corresponding area
of the triangle in the image plane is then used as this triangle's
contribution to the null space fitness measure. Again, this number
should be as low as possible to avoid generating extra images. The
advantage of this method is that it becomes easier to avoid small
undesired images, at the cost of increased computational complexity.

In any case, at this point two numbers describe the fitness of a
genome under study: one describes the overlap of the back-projected
images, the other takes the null space into account. At first, one
might try summing the two fitness values, multiplying each term
with a specific weight. Indeed, our first successful lens 
reconstructions used such a technique. However, using a weighted 
scheme requires the end-user to find and specify acceptable weight 
values. Furthermore, using specific weight values will automatically 
bias the path followed in the search space as the optimization 
routine progresses.

\subsection{A multi-objective genetic algorithm}
			
Fortunately, there is no need to specify a weight factor. Genetic
algorithms are excellent solvers of multi-objective or multi-criterion
optimization problems. Below, some key concepts are introduced. The
interested reader is referred to \citet{Deb} for an in-depth treatment
of this subject.
			
In a multi-objective genetic algorithm a genome has several fitness
measures, each one related to a specific aspect of the optimization
problem. A genome is said to dominate another genome if two criteria
are met: {\em (i)} it is not worse in all fitness measures, and {\em
(ii)} it is strictly better in at least one fitness measure. Using
this concept of dominance, one can identify in a population the
members which are not dominated by any other genome, resulting in the
so-called non-dominated set. The concept of a non-dominated set is
used to devise a new ranking scheme, as there is no longer a single
fitness criterion to base the ranking procedure on. First, the
non-dominated set of the entire population is calculated. The genomes
in this set will receive the highest rank. After removing this set
from the population, a new non-dominated set is identified and these
genomes receive the next-to-highest rank. The procedure is repeated
until all genomes of the population are processed. Afterwards, the
genetic algorithm can proceed as before.

When there are conflicting objectives, there is a whole range of
optimal solutions: the Pareto-optimal front. There exist procedures to
preserve a certain amount of diversity in the population, allowing the
Pareto-optimal front to be well sampled by the genetic algorithm.  In
our specific case however, the objectives are not conflicting and no
further modifications are required.
				
\begin{figure*}
\centering
\subfigure{\includegraphics[origin=c,angle=-90,width=0.32\textwidth]{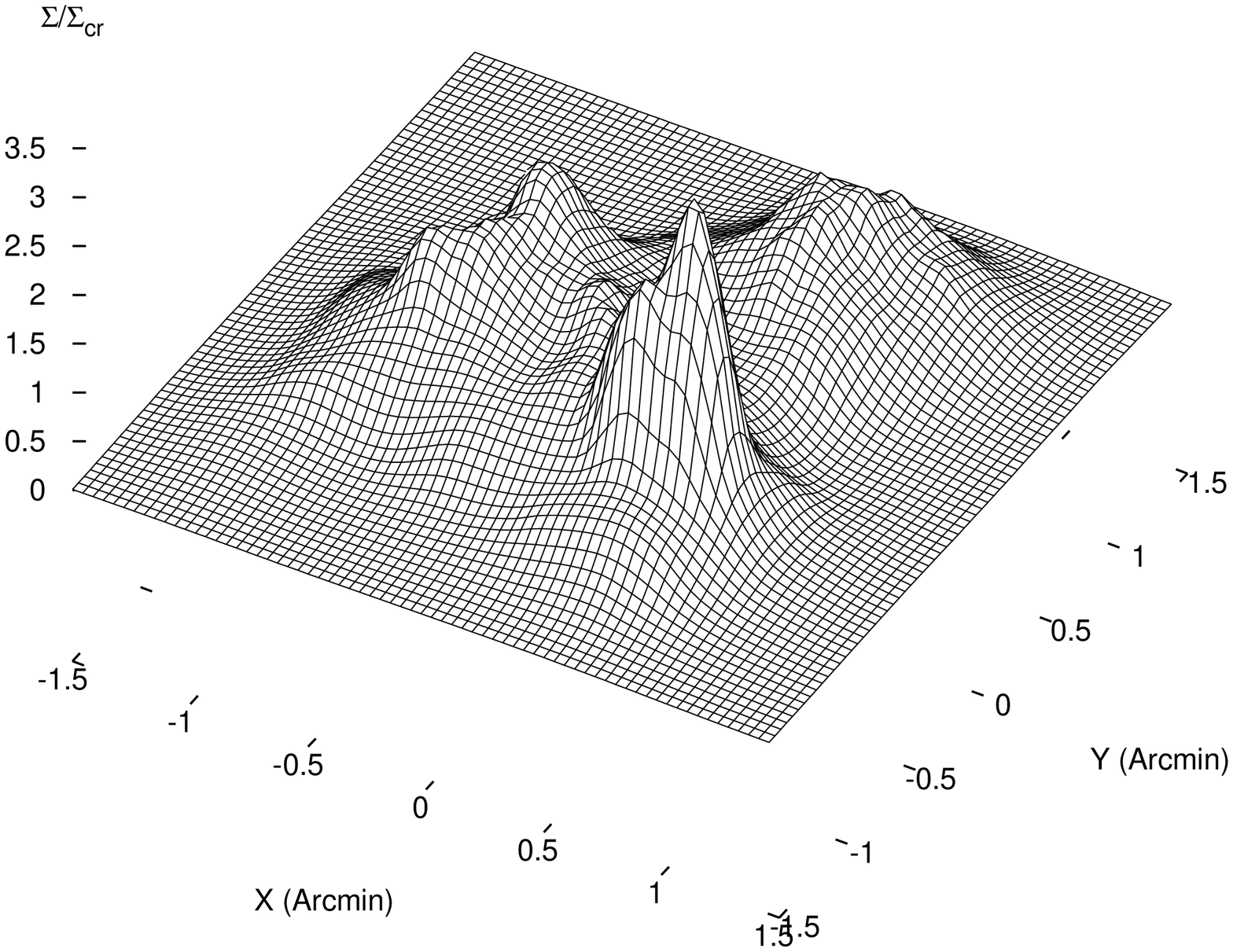}}
\subfigure{\includegraphics[width=0.32\textwidth]{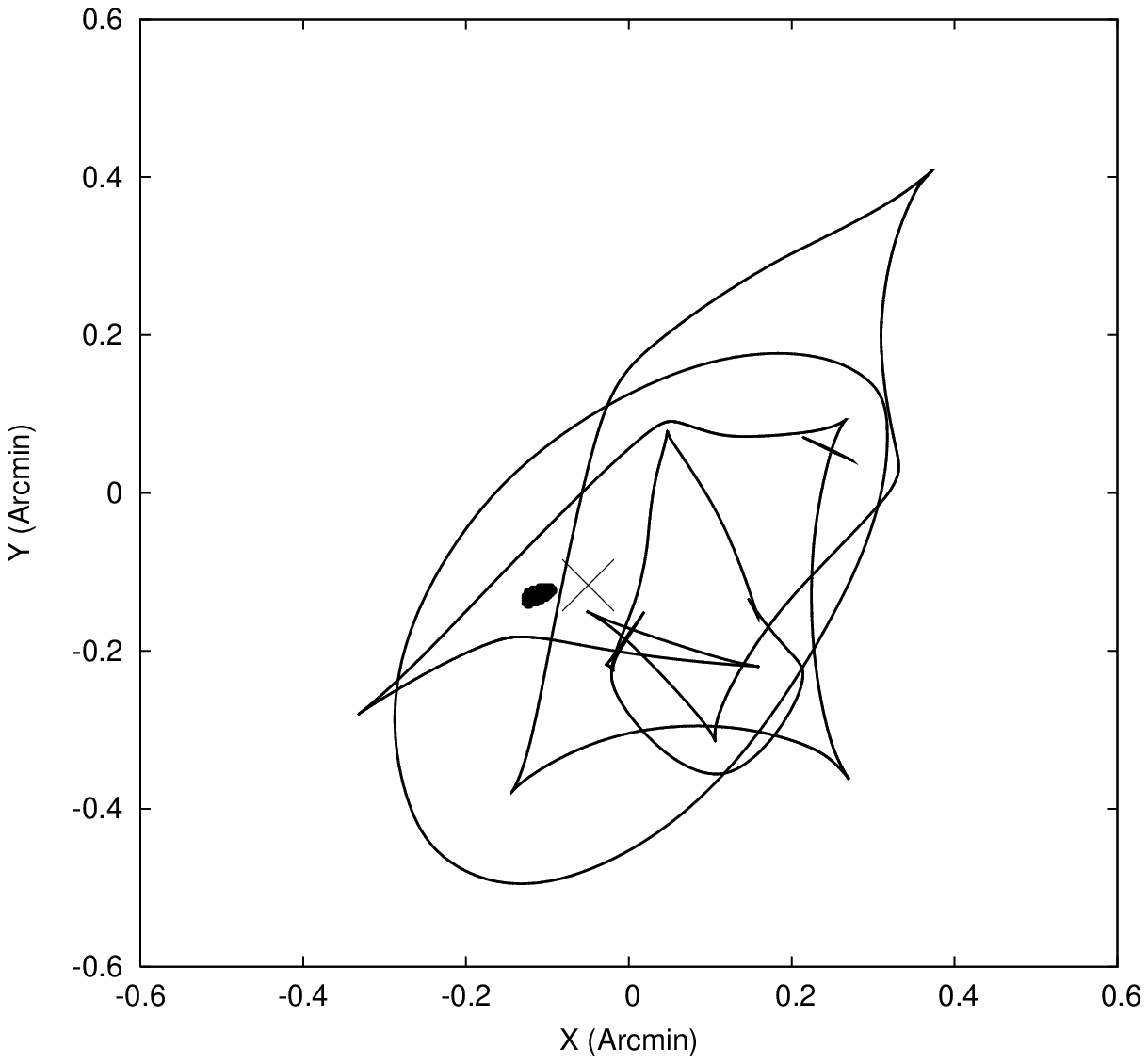}}
\subfigure{\includegraphics[width=0.32\textwidth]{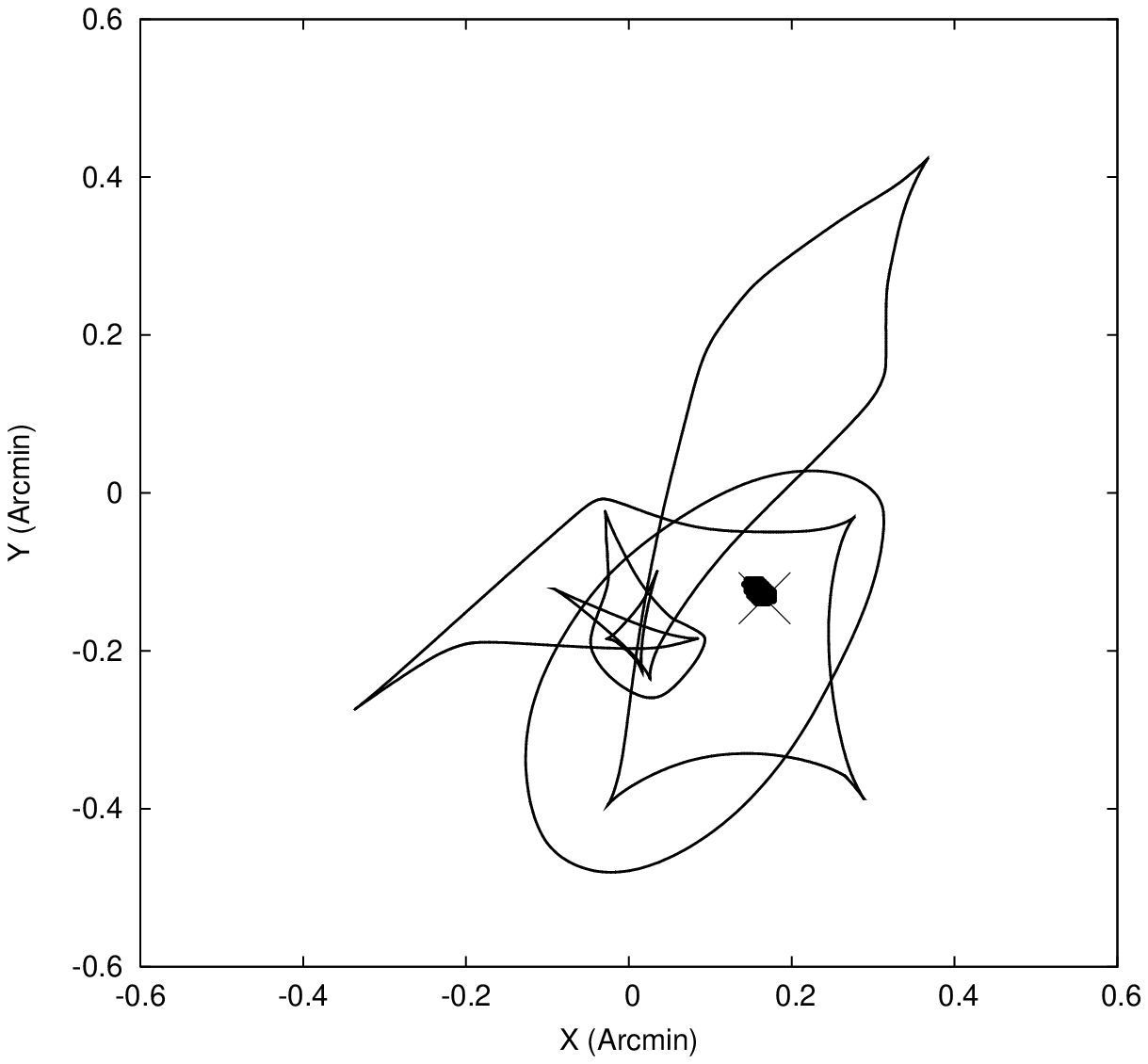}}
\caption{Left panel:~when the null space is taken into account, this
mass distribution was obtained after averaging twenty individual
solutions. Comparing this to the left panel of
Fig.~\ref{fig:reallens} shows that much of the general appearance is
retrieved.  Middle panel and right panel:~the reconstructed sources lie
close to the true source positions (indicated by crosses). 
When these figures are compared to the middle and right panels of 
Fig.~\ref{fig:reallens}, it is clear that much of the same caustic 
structure is present in the reconstruction.}
\label{fig:avglens}
\end{figure*}
		
\section{Simulation results}\label{sec:sim}
			
\begin{figure*}
\centering
\subfigure{\includegraphics[origin=c,width=0.48\textwidth]{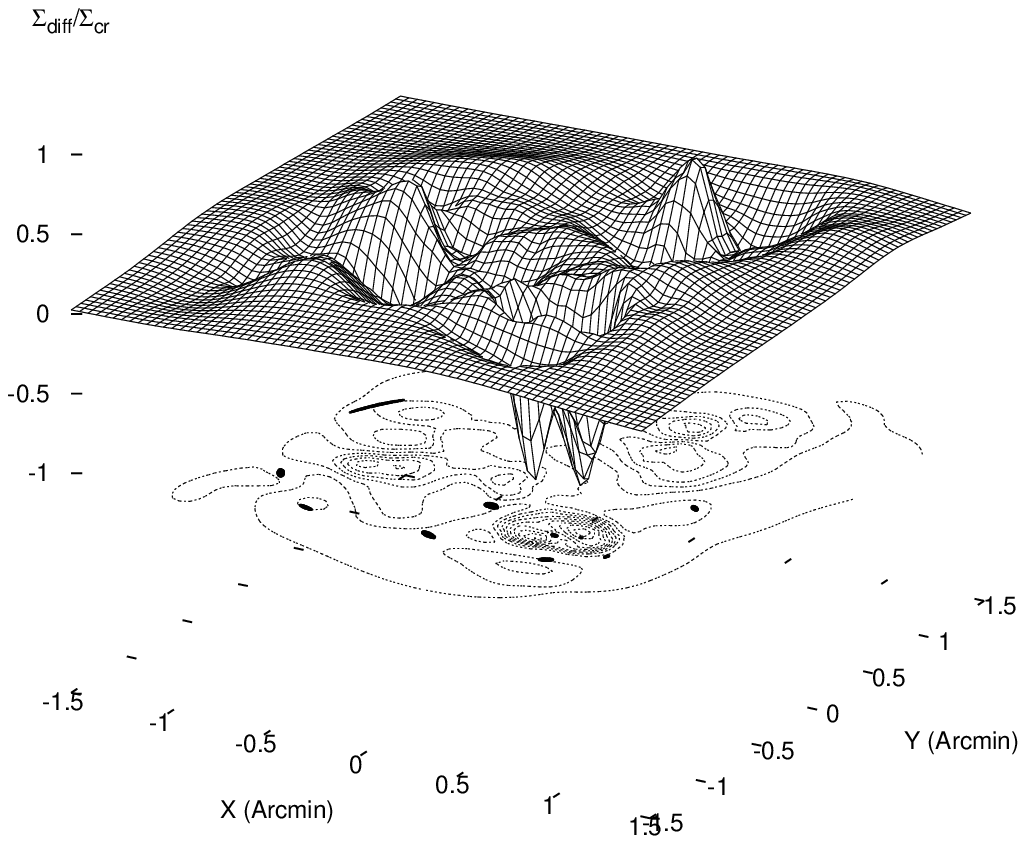}}
\subfigure{\includegraphics[origin=c,width=0.48\textwidth]{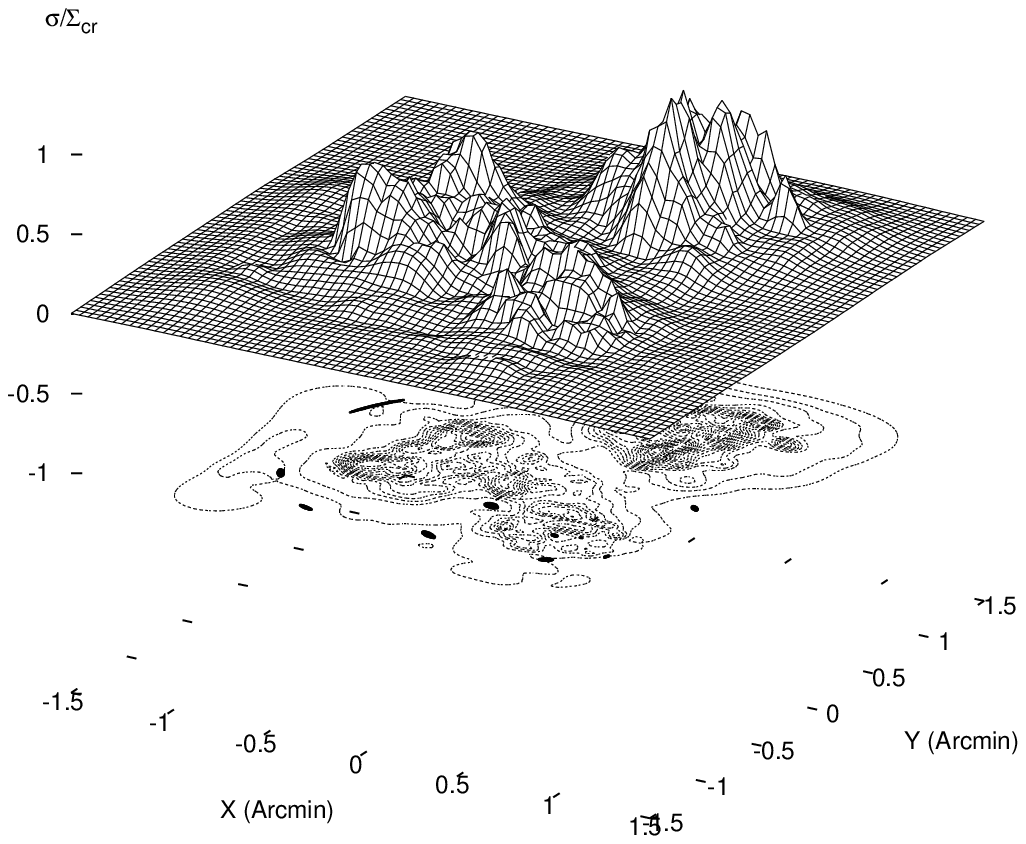}}
\caption{Left panel:~the difference between the
true mass density and the average solution, relative to the critical
mass density for a source at redshift $2.5$. This clearly shows that the true shape
of the mass density peaks cannot be determined accurately. Right
panel:~standard deviation of the twenty individual reconstructed mass
densities. Individual solutions clearly disagree about the exact shape
of the mass density around the peak positions.}
\label{fig:diff3d}
\end{figure*}

\begin{figure*}
\centering
\subfigure{\includegraphics[width=0.48\textwidth]{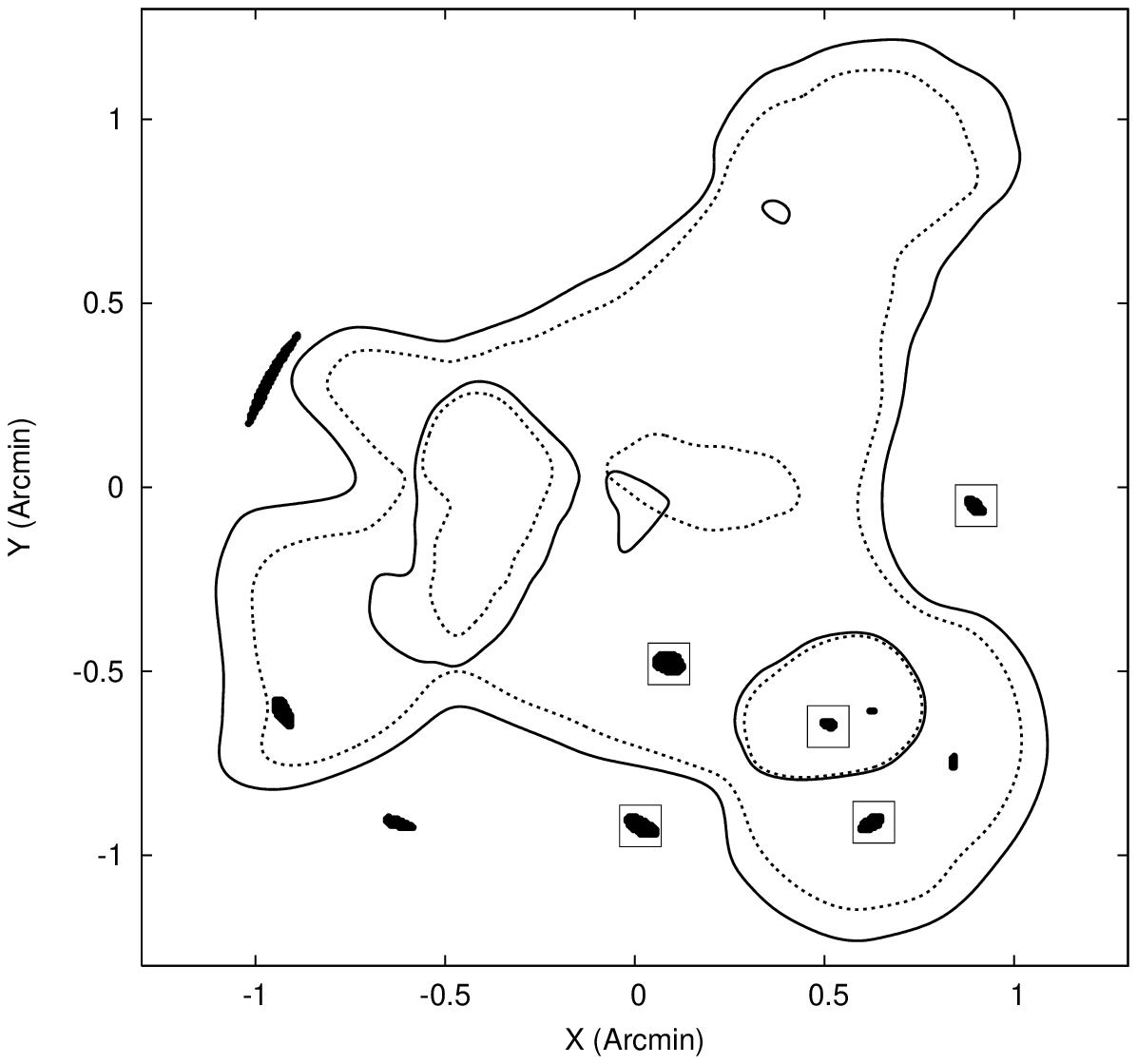}}
\subfigure{\includegraphics[width=0.48\textwidth]{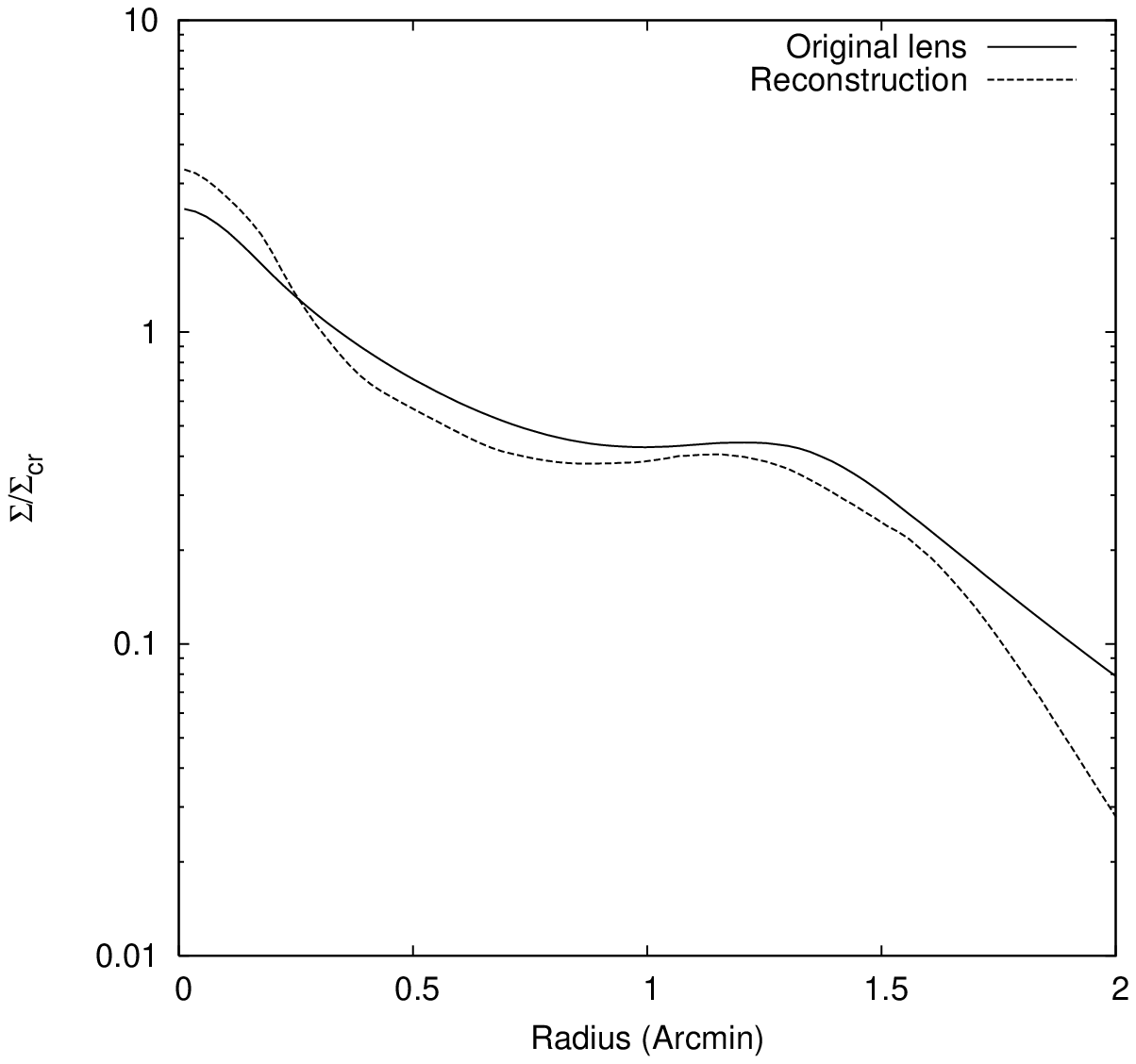}}
\caption{Left panel:~the reconstructed source and lens of
Fig.~\ref{fig:avglens} generated these images and critical lines.  The
correspondence with the left panel of Fig.~\ref{fig:realimages} is
striking. Right panel:~the circularly averaged mass densities of
input lens and reconstruction, as seen from the mass density peak at
$(0.5, -0.5)$.}
\label{fig:avgimages}
\end{figure*}
				
\begin{figure*}
\centering
\subfigure{\includegraphics[width=0.48\textwidth]{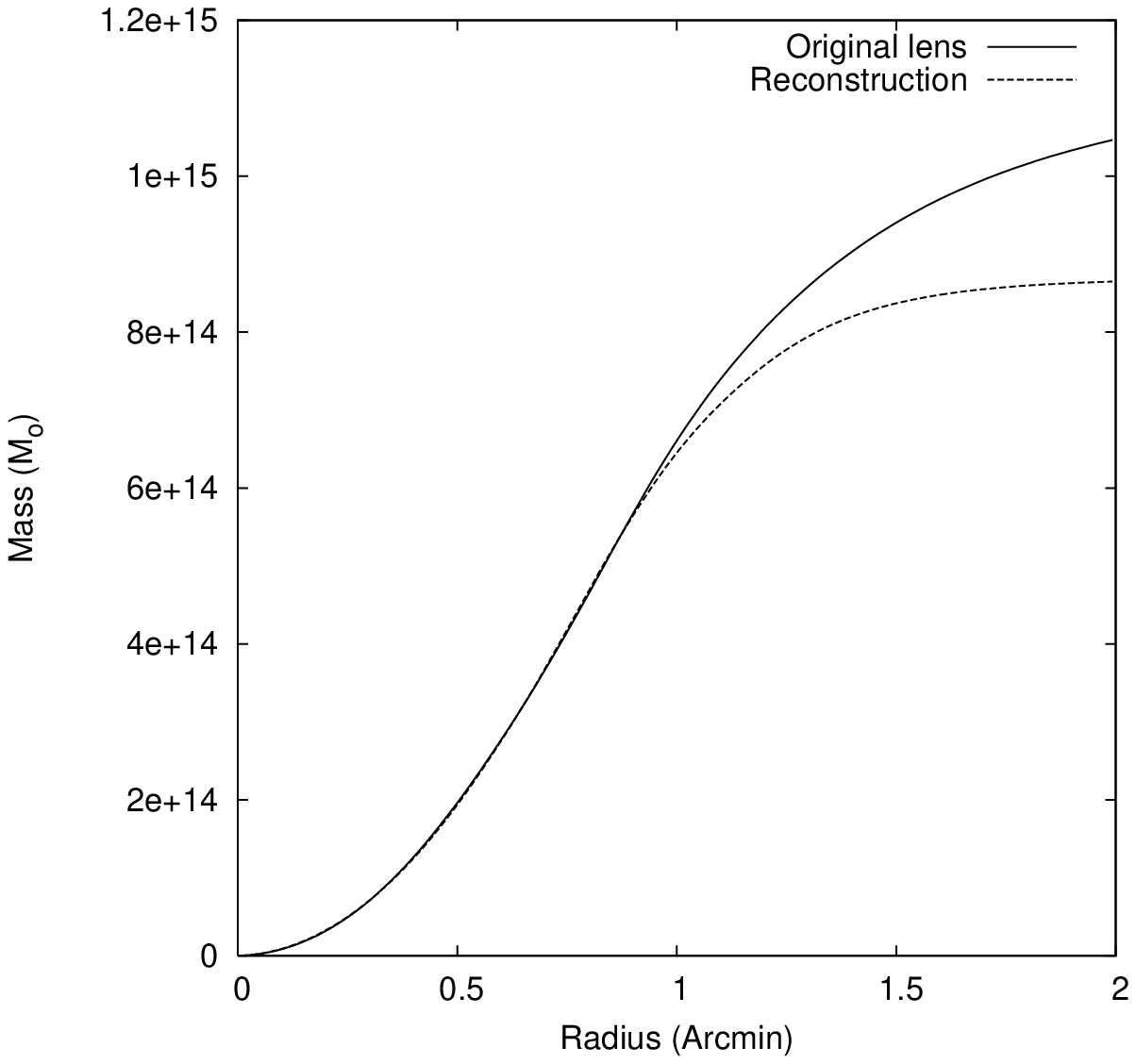}}
\subfigure{\includegraphics[width=0.48\textwidth]{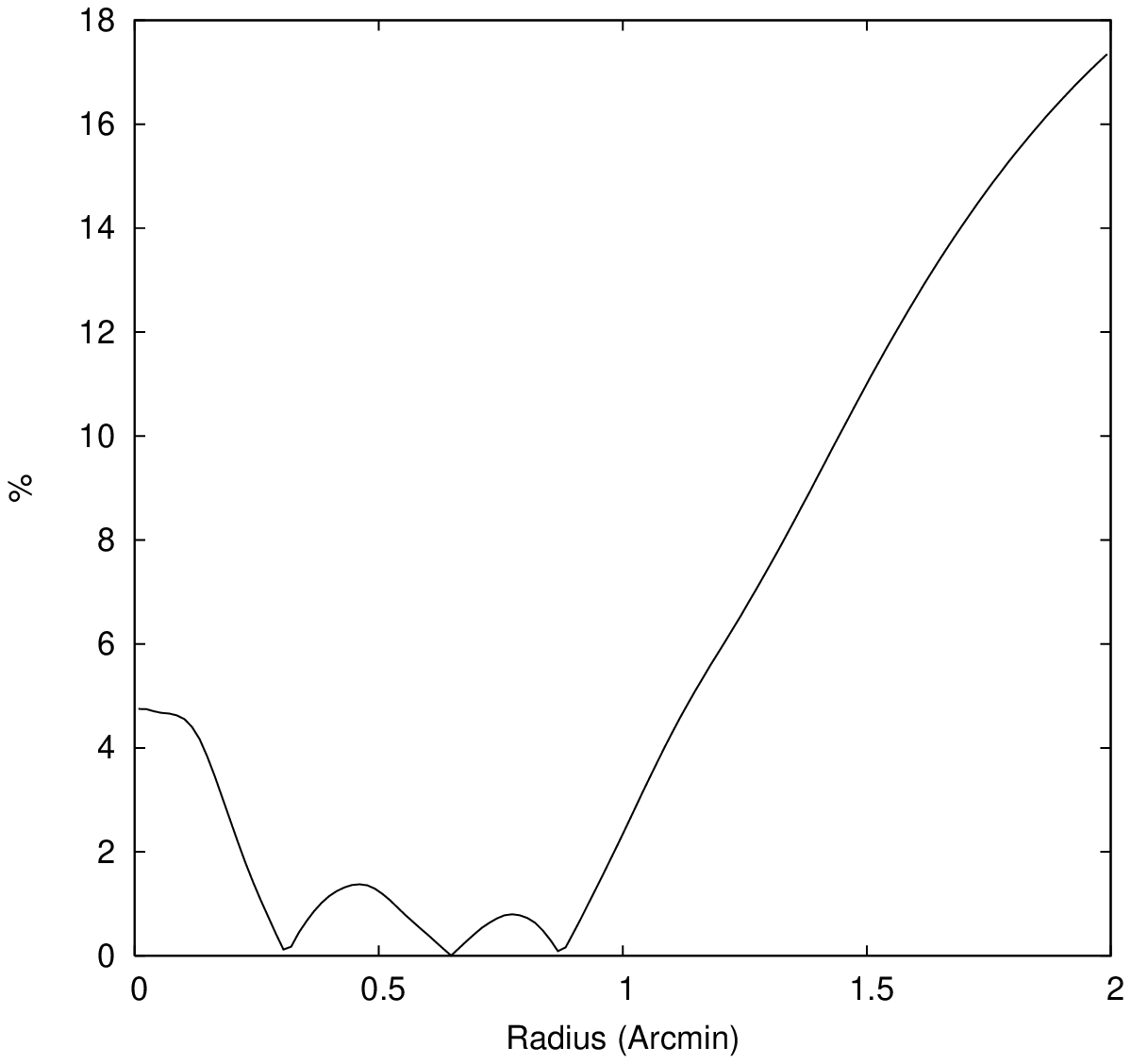}}
\caption{Left panel:~total mass within a specific radius for both input 
lens and reconstruction, as seen from the origin. Clearly, the total mass 
is estimated well within a radius of 1\arcmin. Right panel:~the amount 
of disagreement in total mass within a specific distance from the origin. 
Again, within a radius of 1\arcmin ~the agreement is very good.}
\label{fig:profile}
\end{figure*}

Using the procedure outlined above, it is now straightforward to
generate solutions which indeed only produce the input
images. Averaging twenty such solutions yields the mass density and
reconstructed sources shown in Fig.~\ref{fig:avglens}. Comparing 
this figure with 
Fig.~\ref{fig:reallens}, one immediately notices the striking
resemblance. This proves that our method is capable of reproducing, at
least qualitatively, the mass density distribution of a gravitational
lens based on the positions, redshifts, and shapes of very few images
and on null space information. The peaks in the reconstructed mass
density appear to be somewhat stronger than those of the input
lens. The reconstructed sources are very similar in shape to the true
sources and their positions lie very close to those of the input
sources. The reconstructed sources are, however, more extended than the
input sources. The caustic structure presented in the middle and right
panels of
Fig.~\ref{fig:avglens} are strikingly similar to but more extended than
those of the input lens, presented in the middle and right panels of
Fig.~\ref{fig:reallens}. In this example the method using null space
triangles was used, based on a regular $64\times64$ grid, covering
an area of $3.3\times3.3$ arcmin$^2$.
	
It is interesting to take a look at the difference in mass densities
between original and reconstructed lens. The left panel of
Fig.~\ref{fig:diff3d} shows that the mass density around the peak
positions is not reconstructed very well. Inspecting the right panel
of the same figure, which displays the standard deviation of the
individual solutions, it is clear that precisely these regions differ
strongly among the solutions. When the reconstructed source and lens
are used to reproduce the images, the result shown in the left panel
of Fig.~\ref{fig:avgimages} is obtained. The ten input images are indeed
reproduced and the critical lines closely resemble those of
Fig.~\ref{fig:realimages} (left panel). One cannot ask more 
from any lens inversion algorithm.
The right panel of Fig.~\ref{fig:avgimages}
shows the circularly averaged mass density, centered on the mass
density peak at $(0.5, -0.5)$.  This plot visualizes once more the
overestimated mass density in that region, although the same general
features are clearly present in both original and reconstructed
profile. When the total mass of the lens is calculated, the agreement
is excellent within a radius of 1\arcmin, differing by only a few
percent from the true mass, as is shown in
Fig.~\ref{fig:profile}. Beyond that radius, the difference starts to
increase, indicating that using only the strong lens effect, one
cannot obtain a firm handle on the mass density outside the radius of
the outer images.

It was mentioned before that the reconstructed sources are somewhat
larger (and brighter) than the true sources. Since two sources are
used in this simulation, this cannot be the result of the mass sheet 
degeneracy (e.g. \citet{Saha2000}) and some other type of degeneracy 
must be at work here. In a single-source scenario, the mass-sheet
degeneracy is artificially broken: the mass density of the 
reconstructions quickly drops to zero outside of the grid and 
it is unlikely that the procedure will construct a circular area 
of constant density using (randomly initialized) basis functions 
arranged on a grid.

\section{A second test} \label{sec:sim2}

\begin{figure*}
\subfigure{\includegraphics[width=0.32\textwidth]{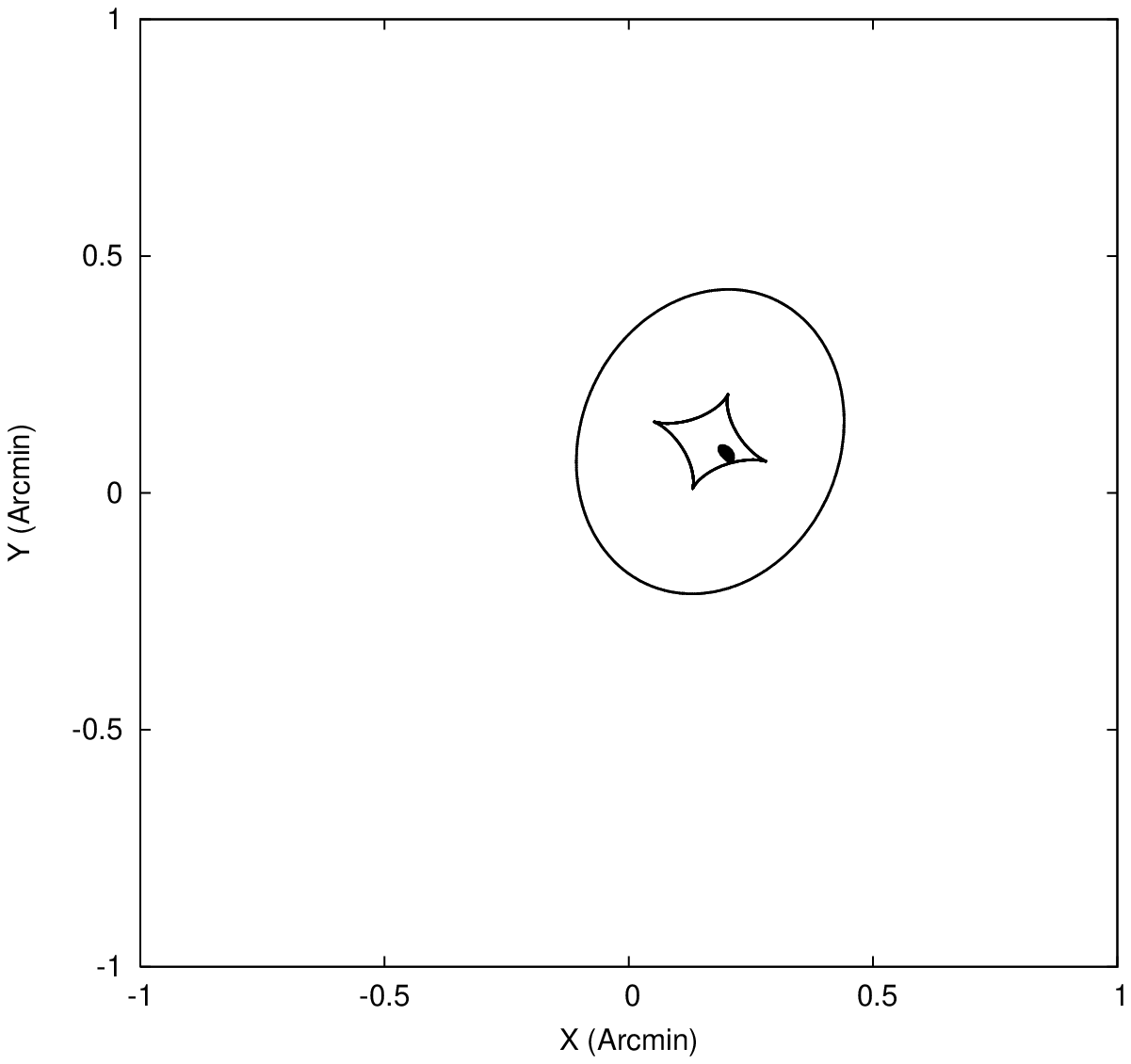}}
\subfigure{\includegraphics[width=0.32\textwidth]{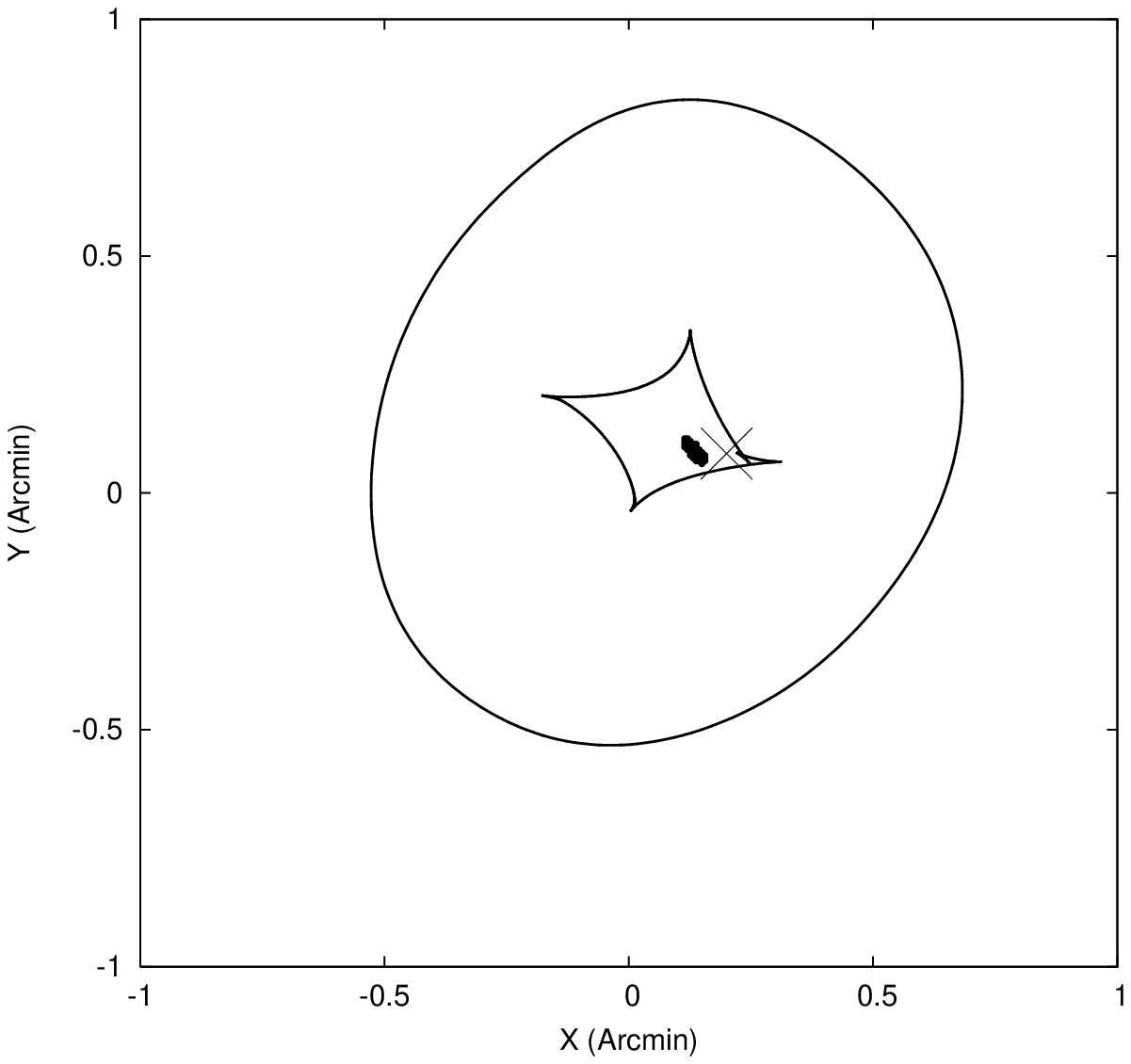}}
\subfigure{\includegraphics[width=0.32\textwidth]{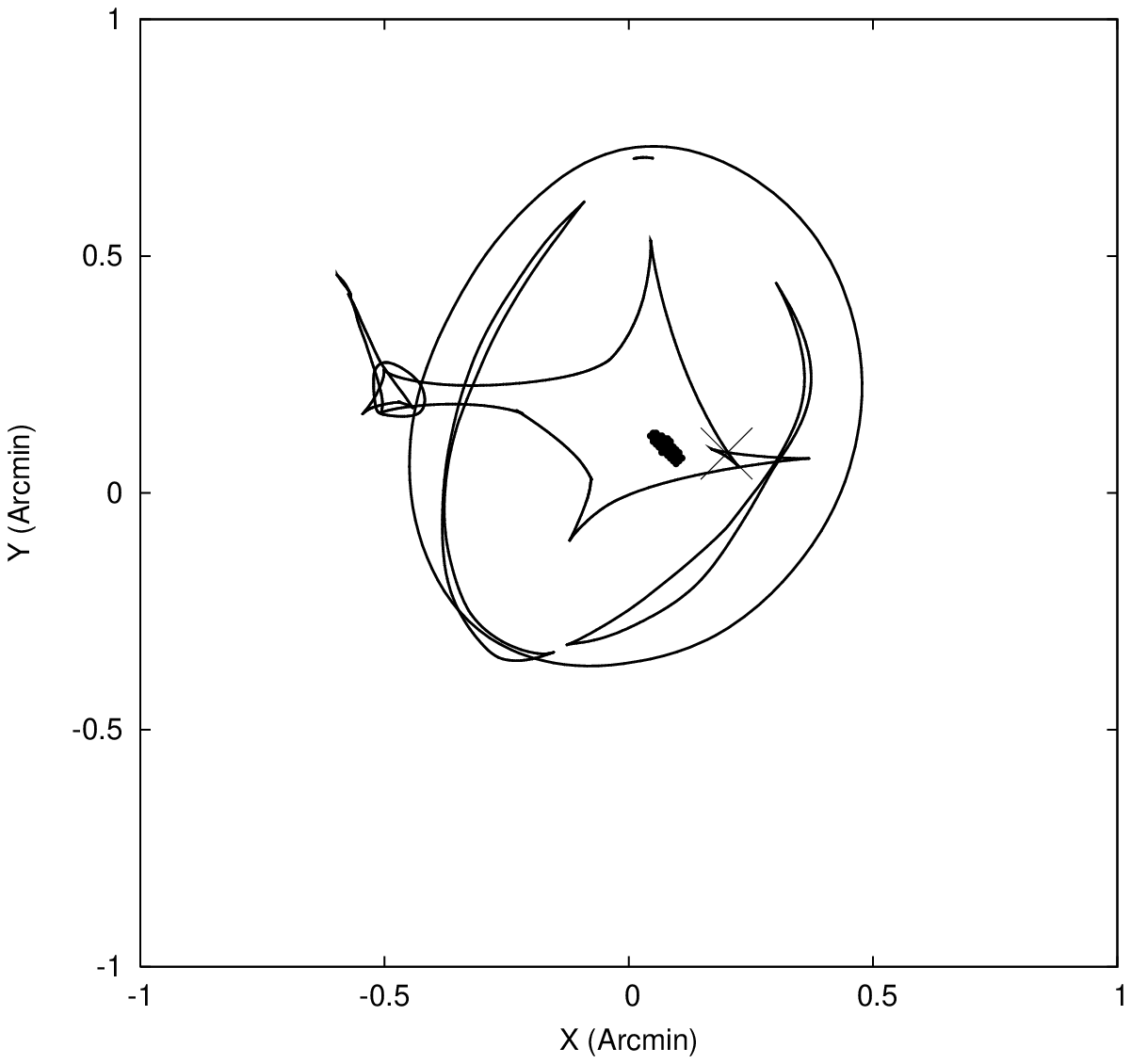}}
\subfigure{\includegraphics[width=0.32\textwidth]{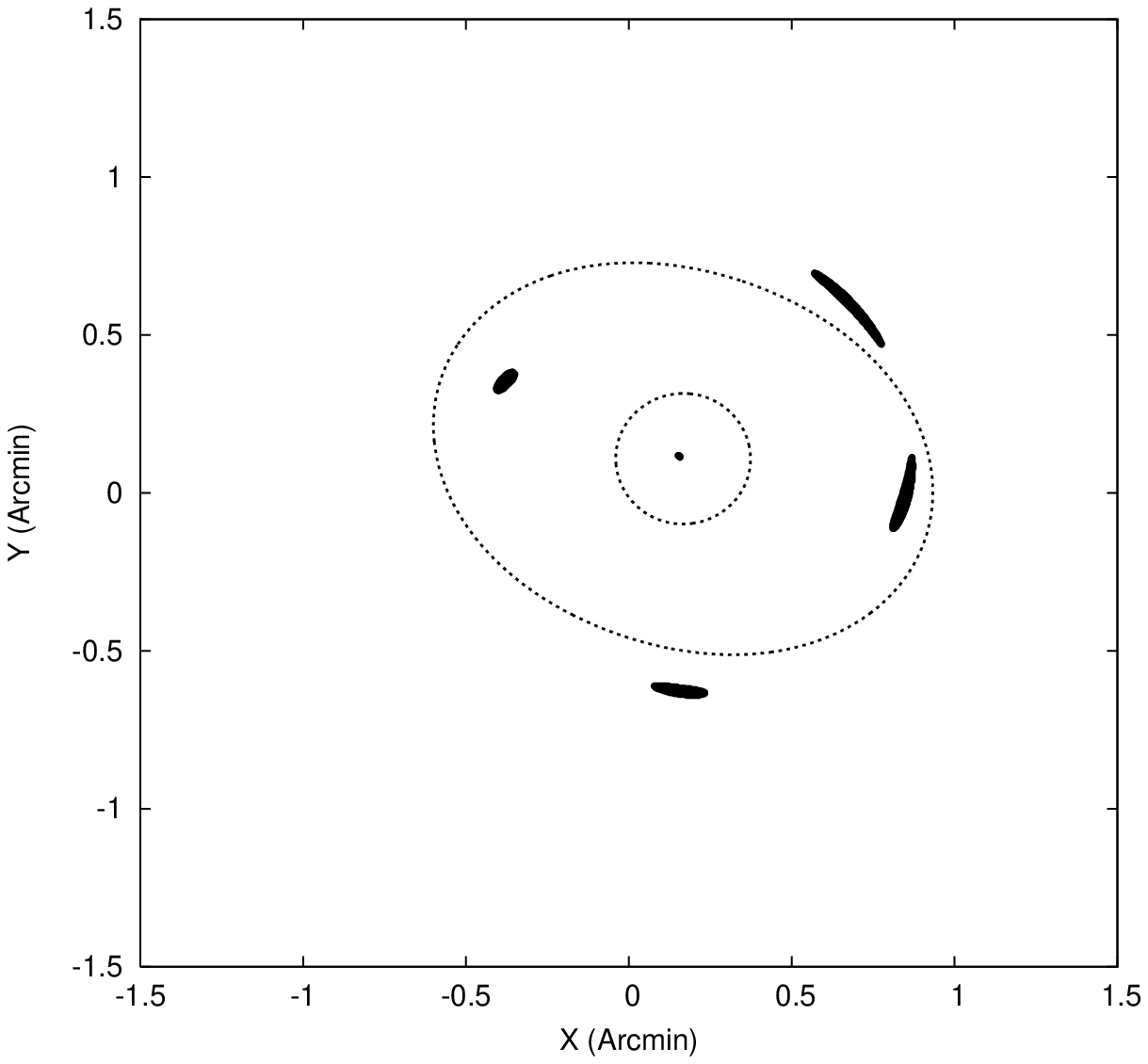}}
\subfigure{\includegraphics[width=0.32\textwidth]{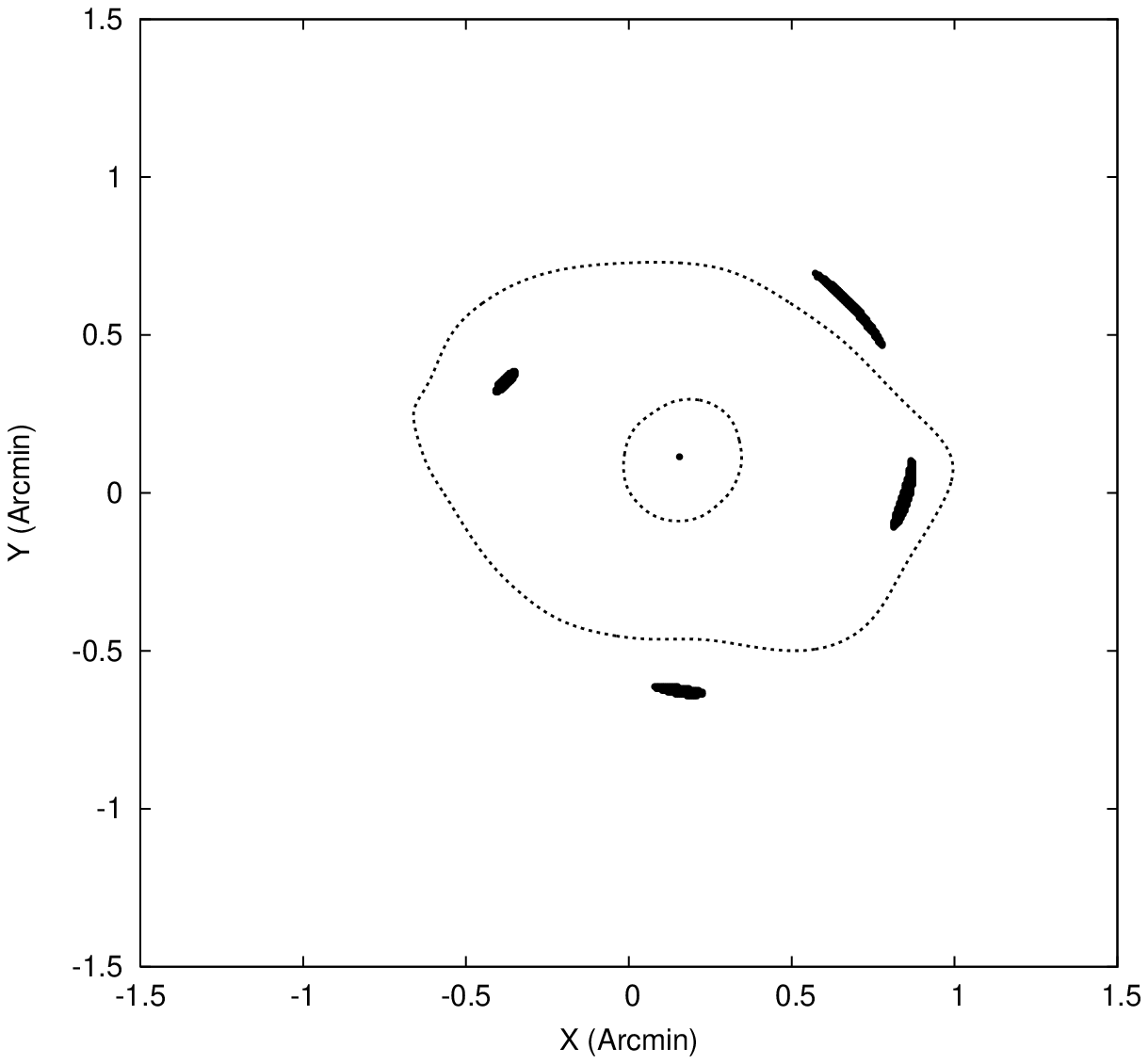}}
\subfigure{\includegraphics[width=0.32\textwidth]{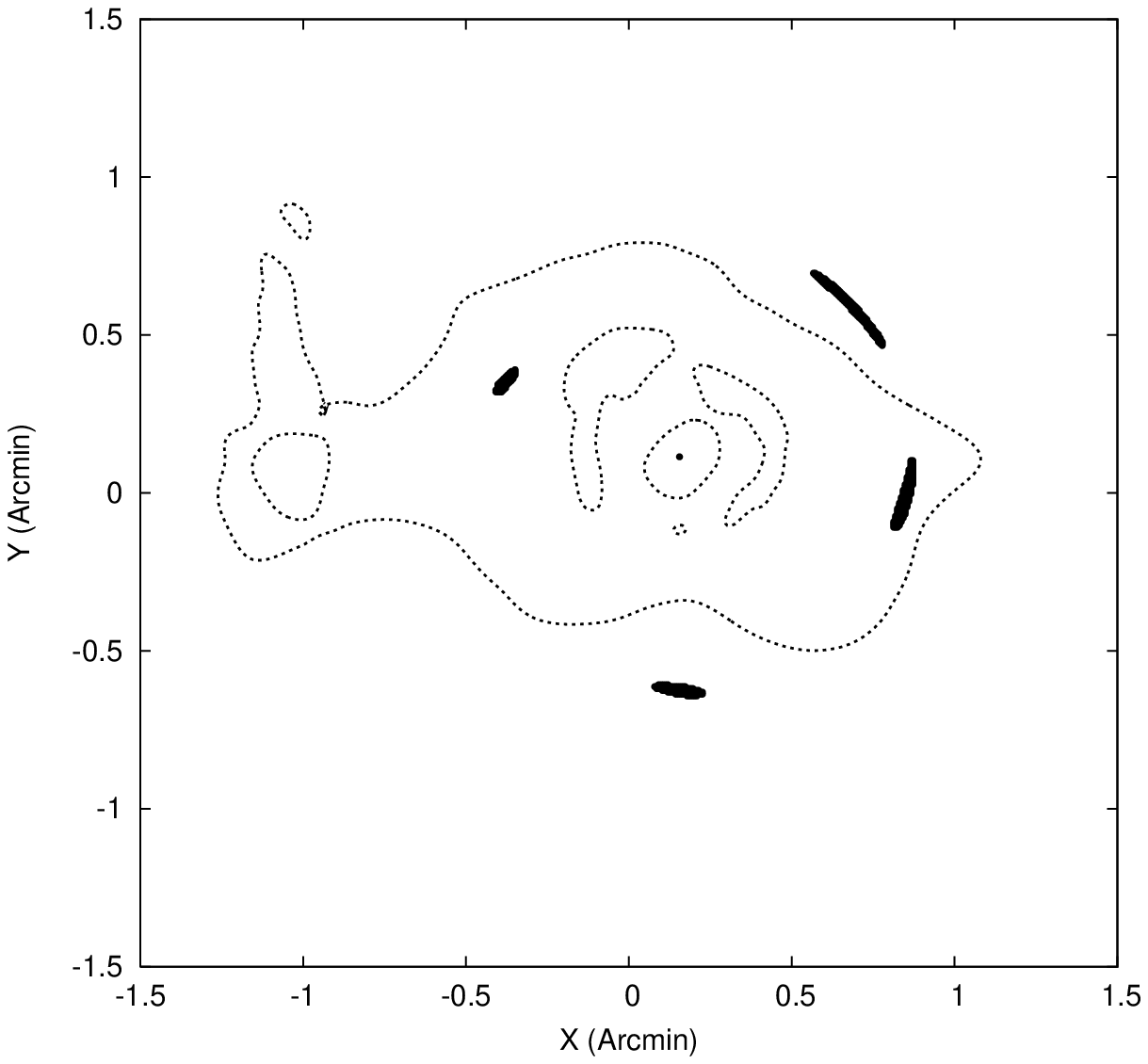}}
\caption{Left panel:~the source used in the simulation of a 
gravitational lens system with an elliptical mass distribution, 
relative to the caustic structure. This situation creates five images,
which are used as the input of the inversion routine. 
Center panel:~the reconstructed source, relative to the predicted
caustics when using a moderately subdivided grid and averaging ten
individual solutions. Note that the source and the caustic structure are 
more extended than their original counterparts. Shown below are the 
five resulting images as well as the critical lines. 
Right panel:~similar to the center panel, but using a finely subdivided 
grid. The fitness measures did not improve considerably compared to the
situation in the center panel.}
\label{fig:ell}
\end{figure*}

One might argue that a system with two sources and ten images still 
provides a relatively large number of constraints. For this reason, 
let us now turn to a simple five-image system, created by an 
elliptical mass distribution. The source and its corresponding images 
are depicted in the left panel of Fig.~\ref{fig:ell}. The redshifts of 
the source and the lens are $2.5$ and $0.45$ respectively. 
Note that in this situation the mass-sheet degeneracy is artificially
broken, as explained above.

In this case, the inversion algorithm seems to easily produce
solutions causing a critical line to intersect the two
rightmost images. To avoid this, the brightness overlap and
positional overlap were used as two separate fitness measures. 
Since their combination can be seen as a weighting scheme, 
decoupling them allows a broader search in the model space.
Afterwards, the genetic algorithm was indeed able to consistently 
find solutions which reproduce the input images. After averaging
ten solutions obtained using a moderately subdivided grid,
the results shown in the center panel of Fig.~\ref{fig:ell} 
were obtained. This clearly resembles the true situation, although
the source is again larger and brighter. If the grid is
subdivided further, this results in the situation depicted
in the right panel of Fig.~\ref{fig:ell}, showing more complex 
critical lines and corresponding caustic structure. 

It is important to note that the resulting genome fitness did not 
improve significantly by using a finer grid. In practice,
due to a lack of constraints, one may simply opt for the least 
complex reconstruction. However,
as is shown in the right panel of Fig.~\ref{fig:ell}, the algorithm
is indeed able to produce viable solutions containing small-scale
variability introduced by using approximately one thousand basis
functions.
This ability to handle different scales well is, of course, required
to qualify as a truly non-parametric technique.

\section{Discussion and conclusion} \label{sec:conc}

In this article we have presented an extension of our previously
proposed inversion algorithm, which was intended to be used on a
strong lensing system with many multiply imaged sources. Since
currently the number of known strong lensing systems with few sources
far exceed those with many, we felt it was appropriate to
study the necessary extensions of the original procedure to allow such
systems to be inverted non-parametrically. The modifications
presented above still use only physically motivated arguments and do
not impose any prior on the reconstructed mass density. The inclusion
of the null space led to an additional fitness measure, which can be
handled efficiently and without the need of a weight parameter using a
multi-objective genetic algorithm. In a similar way one could
incorporate even more constraints, e.g. from the weak lensing regime
or from time delay measurements, without the need for weights.
This generic way of handling various types of constraints is clearly
a great advantage of the proposed inversion method.

We found that including brightness information is in this case
necessary to ascertain acceptable source reconstructions.  It is
interesting to note that even without this extra information, the
inclusion of the null space already brings out most of the general
structure of the mass distribution. This suggests that uncertainty in
measured image intensities will not prevent a good reconstruction of
the mass density, which is important from a practical point of
view. 		

Although the examples shown
above seem to suggest a bias towards larger and brighter reconstructed
sources, it is easy to create a scenario in which the genetic algorithm 
will cause the reconstructed sources to be smaller and dimmer. This
kind of degeneracy can only be resolved if more data are available or
some prior is imposed on the source.
The fact that the true source features cannot be obtained with certainty 
even when all the information is available with the greatest accuracy,
indicates that care should be taken when interpreting source data
obtained by strong lens inversion in general. 

Deviations from the true mass density are to be expected, not only
because of the sparse sampling of the lens equation, but also because
of the inherent degenerate nature of the lens inversion
problem. Nevertheless, the similarities between the original lens
systems and
the reconstructions are remarkable, especially when taking into
consideration that no prior information about the lens was
needed. This indicates that much information is present in a strong
lensing system and that it can be extracted non-parametrically using
only information about the observed images.

Above, we have investigated only the ideal situation, when all image 
positions and fluxes are known accurately. Practical difficulties
like incomplete null space information, obscured images, measurement
errors or the effect of a point spread function will further reduce
the information content of a strong lensing system. The precise
impact of these factors will require further research. Furthermore,
if relatively large images are present, it is likely that the simple
technique to determine the overlap between images will fail. We are
currently exploring methods to take image substructure into account
to be able to handle such cases effectively.

\section*{Acknowledgment}

We would like to thank the anonymous referee for thoroughly
reviewing the manuscript and supplying many valuable suggestions.

\bsp 
\label{lastpage}


\begin{thebibliography}{19}
\expandafter\ifx\csname natexlab\endcsname\relax\def\natexlab#1{#1}\fi

\bibitem[{{Abdelsalam} {et~al.}(1998){Abdelsalam}, {Saha}, \&
  {Williams}}]{Abdelsalam}
{Abdelsalam} H.~M., {Saha} P., {Williams} L.~L.~R., 1998, \mnras, 294, 734

\bibitem[{{Blandford} \& {Kochanek}(1987)}]{Blandford}
{Blandford} R.~D., {Kochanek} C.~S., 1987, \apj, 321, 658

\bibitem[{{Brada{\v c}} {et~al.}(2005){Brada{\v c}}, {Schneider}, {Lombardi},
  \& {Erben}}]{Bradac}
{Brada{\v c}} M., {Schneider} P., {Lombardi} M., {Erben} T., 2005, \aap, 437,
  39

\bibitem[{{Broadhurst} {et~al.}(2005){Broadhurst}, {Ben{\'{\i}}tez}, {Coe},
  {Sharon}, {Zekser}, {White}, {Ford}, {Bouwens}, {Blakeslee}, {Clampin},
  {Cross}, {Franx}, {Frye}, {Hartig}, {Illingworth}, {Infante}, {Menanteau},
  {Meurer}, {Postman}, {Ardila}, {Bartko}, {Brown}, {Burrows}, {Cheng},
  {Feldman}, {Golimowski}, {Goto}, {Gronwall}, {Herranz}, {Holden}, {Homeier},
  {Krist}, {Lesser}, {Martel}, {Miley}, {Rosati}, {Sirianni}, {Sparks},
  {Steindling}, {Tran}, {Tsvetanov}, \& {Zheng}}]{Broadhurst}
  {Broadhurst} T., {Ben{\'{\i}}tez} N., {Coe} D., {Sharon} K., {Zekser} K.,
  {White} R., {Ford} H., {Bouwens} R., {Blakeslee} J.,et al., 
   2005, \apj, 621, 53

\bibitem[{{Charbonneau}(1995)}]{Charbonneau}
{Charbonneau} P., 1995, \apjs, 101, 309

\bibitem[{Deb(2001)}]{Deb}
Deb K., 2001, Multi-Objective Optimization Using Evolutionary Algorithms. John
  Wiley \& Sons, Inc., New York, NY, USA

\bibitem[{Diego {et~al.}(2005)Diego, Protopapas, Sandvik, \& Tegmark}]{Diego}
Diego J.~M., Protopapas P., Sandvik H.~B., Tegmark M., 2005, \mnras, 360, 477

\bibitem[{Holland(1975)}]{Holland}
Holland J.~H., 1975, Adaptation in natural and artificial systems. University
  of Michigan Press, Ann Arbor

\bibitem[{{Keeton}(2001)}]{KeetonGravlens}
{Keeton} C.~R., 2001, ArXiv Astrophysics e-prints

\bibitem[{{Kneib} {et~al.}(1993){Kneib}, {Mellier}, {Fort}, \&
  {Mathez}}]{KneibLenstool}
{Kneib} J.~P., {Mellier} Y., {Fort} B., {Mathez} G., 1993, \aap, 273, 367

\bibitem[{{Kochanek} \& {Narayan}(1992)}]{LensCLEAN}
{Kochanek} C.~S., {Narayan} R., 1992, \apj, 401, 461

\bibitem[{{Koopmans}(2005)}]{Koopmans}
{Koopmans} L.~V.~E., 2005, \mnras, 363, 1136

\bibitem[{Koza(1992)}]{Koza}
Koza J.~R., 1992, Genetic Programming: On the Programming of Computers by Means
  of Natural Selection (Complex Adaptive Systems). {The MIT Press}

\bibitem[{{Liesenborgs} {et~al.}(2006){Liesenborgs}, {De Rijcke}, \&
  {Dejonghe}}]{Liesenborgs}
{Liesenborgs} J., {De Rijcke} S., {Dejonghe} H., 2006, \mnras, 367, 1209

\bibitem[{{Plummer}(1911)}]{Plummer}
{Plummer} H.~C., 1911, \mnras, 71, 460

\bibitem[{{Saha}(2000)}]{Saha2000}
{Saha} P., 2000, \aj, 120, 1654

\bibitem[{{Saha} \& {Williams}(1997)}]{Saha}
{Saha} P., {Williams} L.~L.~R., 1997, \mnras, 292, 148

\bibitem[{{Schneider} {et~al.}(1992){Schneider}, {Ehlers}, \&
  {Falco}}]{SchneiderBook}
{Schneider} P., {Ehlers} J., {Falco} E.~E., 1992, {Gravitational Lenses}.
  Gravitational Lenses, XIV, 560 pp.~112 figs..~Springer-Verlag Berlin
  Heidelberg New York.~ Also Astronomy and Astrophysics Library

\bibitem[{{Shamos}(1979)}]{Shamos}
{Shamos} M.~I., 1979, PhD thesis, Yale University

\end{thebibliography}
\end{document}